\documentclass[12pt,letterpaper]{article}

\usepackage{natbib}
\usepackage[hidelinks]{hyperref}
\usepackage[ left=1in, top=1in, right=1in, bottom=1in]{geometry}
\usepackage{graphicx,bm,colonequals,amsmath,amssymb,url}
\usepackage{array,tabularx,multirow}
\usepackage{enumitem,algpseudocode}
\usepackage[font={footnotesize}]{caption,subcaption}

\usepackage{titlesec}
\titlespacing*\subsubsection{0pt}{12pt plus 4pt minus 2pt}{2pt plus 2pt minus 2pt}

\bibpunct[, ]{(}{)}{;}{a}{,}{,}

\usepackage{amsthm}
\newtheoremstyle{propstyle} 
    {2mm}                    
    {1mm}                    
    {\itshape}                   
    {}                           
    {\scshape}                   
    {.}                          
    {.5em}                       
    {}  
\theoremstyle{propstyle}
\newtheorem{prop}{Proposition}

\newcommand{\bb}{\mathbf{b}}
\newcommand{\bs}{\mathbf{s}}

\newcommand{\by}{\mathbf{y}}

\newcommand{\bA}{\mathbf{A}}

\newcommand{\bW}{\mathbf{W}}

\newcommand{\bL}{\mathbf{L}}
\newcommand{\bI}{\mathbf{I}}

\newcommand{\bU}{\mathbf{U}}
\newcommand{\bV}{\mathbf{V}}
\newcommand{\bK}{\mathbf{K}}

\newcommand{\bB}{\mathbf{B}}
\newcommand{\bC}{\mathbf{C}}

\newcommand{\bfzero}{\mathbf{0}}

\newcommand{\bfmu}{\bm{\mu}}

\newcommand{\bftheta}{\bm{\theta}}
\newcommand{\bfeta}{\bm{\eta}}
\newcommand{\bfnu}{\bm{\nu}}
\newcommand{\bfdelta}{\bm{\delta}}

\newcommand{\bftau}{\bm{\tau}}
\newcommand{\bfomega}{\bm{\omega}}

\newcommand{\bfSigma}{\bm{\Sigma}}

\newcommand{\bfPsi}{\bm{\Psi}}

\newcommand{\im}{{i_1,\ldots,i_m}}
\newcommand{\jm}{{j_1,\ldots,j_m}}
\newcommand{\jmp}{{j_1,\ldots,j_{m+1}}}
\newcommand{\jmm}{{j_1,\ldots,j_{m-1}}}
\newcommand{\jk}{{j_1,\ldots,j_k}}
\newcommand{\jl}{{j_1,\ldots,j_l}}
\newcommand{\jlp}{{j_1,\ldots,j_{l+1}}}
\newcommand{\jM}{{j_1,\ldots,j_M}}
\newcommand{\etaset}{\mathcal{E}}

\newcommand{\var}{var}
\newcommand{\cov}{cov}

\newcommand{\GP}{GP}

\newcommand{\blockdiag}{blockdiag}

\newcommand{\knots}{\mathcal{Q}}
\newcommand{\node}{\mathcal{N}}
\newcommand{\order}{\mathcal{O}}

\newcommand{\domain}{\mathcal{D}}
\newcommand{\pp}{\tau}
\newcommand{\locs}{\mathcal{S}}

\newcommand{\dg}{\mbox{$^{\circ}$}}

\title{A multi-resolution approximation for massive spatial datasets}

\author{Matthias Katzfuss\thanks{Department of Statistics, Texas A\&M University. \texttt{katzfuss@gmail.com}}}

\date{Published in Journal of the American Statistical Association (DOI: \href{http://dx.doi.org/10.1080/01621459.2015.1123632}{10.1080/01621459.2015.1123632})}

\begin{document}

\maketitle

\begin{abstract}

Automated sensing instruments on satellites and aircraft have enabled the collection of massive amounts of high-resolution observations of spatial fields over large spatial regions. If these datasets can be efficiently exploited, they can provide new insights on a wide variety of issues. However, traditional spatial-statistical techniques such as kriging are not computationally feasible for big datasets. 
We propose a multi-resolution approximation ($M$-RA) of Gaussian processes observed at irregular locations in space. The $M$-RA process is specified as a linear combination of basis functions at multiple levels of spatial resolution, which can capture spatial structure from very fine to very large scales. The basis functions are automatically chosen to approximate a given covariance function, which can be nonstationary. All computations involving the $M$-RA, including parameter inference and prediction, are highly scalable for massive datasets. Crucially, the inference algorithms can also be parallelized to take full advantage of large distributed-memory computing environments. In comparisons using simulated data and a large satellite dataset, the $M$-RA outperforms a related state-of-the-art method.

\end{abstract}

\subsection*{Keywords}
Basis functions; Distributed computing; Full-scale approximation; Gaussian process; Kriging; Satellite data.

\section{Introduction \label{sec:intro}}

Automated sensing instruments on satellites and aircraft have enabled the collection of massive amounts of high-resolution observations of spatial fields over large and inhomogenous spatial domains. If these kinds of datasets can be efficiently exploited, they can provide new insights on a wide variety of issues, such as greenhouse gas concentrations for climate change, soil properties for precision agriculture, and atmospheric states for weather forecasting. Based (implicitly or explicitly) on Gaussian processes, the field of spatial statistics provides a rich toolkit for the analysis of such data, including estimating unknown parameters, predicting the spatial field at unobserved locations, and properly quantifying uncertainty in the predictions and parameters \citep[e.g.,][]{Cressie2011}. 

However, traditional spatial-statistical techniques such as kriging are not computationally feasible for big datasets, because dense $n \times n$ matrices need to be decomposed, where $n$ is the number of measurements. General-purpose methods such as the preconditioned conjugate gradient algorithm \citep[e.g.,][]{Golub2012} or probabilistic projections \citep[e.g.,][]{Halko2011,Stein2013} can solve large linear systems. But it is challenging for such algorithms to calculate the determinant of the data covariance matrix also required for likelihood-based inference, as this covariance is huge, often dense, and might have a slowly decaying spectrum.


More specialized methods for approximating spatial inference that explicitly try to exploit spatial information in the data have been proposed in the literature, but most of them either require restrictive assumptions about the covariance function \citep[e.g.,][]{Fuentes2007,Lindgren2011a}, or they ignore much of the fine-scale dependence \citep[e.g.,][]{Higdon1998,Mardia1998,Calder2007,Cressie2008,Katzfuss2009,Lemos2009,Katzfuss2010,Katzfuss2011} or the large-scale dependence \citep[e.g.,][]{Furrer2006,Kaufman2008,Shaby2010}. Composite-likelihood methods \citep[e.g.,][]{Vecchia1988,Curriero1999,Stein2004,Caragea2007,Bevilacqua2012,Eidsvik2012} achieve computational feasibility by treating (blocks of) observations as conditionally independent, but it is not clear how to obtain proper joint predictive distributions for locations in different blocks.


We propose here a multi-resolution approximation ($M$-RA) of Gaussian processes observed at irregular (i.e., non-gridded) locations in space. The $M$-RA process is specified as a linear combination of basis functions at multiple levels of spatial resolution, which can capture spatial structure from very fine to very large scales. Multi-resolution models \citep[e.g.][]{Chui1992,Johannesson2007,Nychka2012} 
have been very successful in spatial statistics, due to their ability to flexibly capture dependence at multiple spatial scales while being computationally feasible. In constrast to these existing methods, in our $M$-RA the basis functions and the distributions of their weights are chosen to ``optimally'' approximate a given covariance function, without requiring any restrictions on this covariance function. The basis functions in each region at each resolution are chosen iteratively according to the rules of the predictive process \citep{Quinonero-Candela2005,Banerjee2008}, based on a recursive partitioning of the spatial domain into smaller and smaller subregions, and a set of ``knot'' locations in each region. The $M$-RA is a generalization of the full-scale approximation \citep{Snelson2007,Sang2011a,Sang2012}, a current state-of-the-art method for covariance approximations for large spatial data, which has only one level of domain partitioning and one resolution of basis functions. We will compare the two approaches extensively.


Inference for basis-function models essentially consists of obtaining the posterior distribution of the basis-function weights. In previous approaches, computationally feasible inference has been achieved by limiting the number of basis functions to be small \citep[e.g.,][]{Higdon1998,Quinonero-Candela2005,Cressie2008} or by specifying the precision matrix of the weight vector to be diagonal or sparse \citep[e.g.,][]{Higdon1998,Lindgren2011a,Nychka2012}. The $M$-RA combines both approaches: It results in a multi-resolution (block) sparse precision matrix, but the number of spatial basis functions within each region is small, allowing repeated application of the Sherman-Morrison-Woodbury formula \citep{Sherman1950,Woodbury1950}. This leads to a highly scalable inference algorithm for the $M$-RA (which could also be applied to any multi-resolution basis-function model with a similar structure). Crucially, based on previous work \citep{Katzfuss2014} describing parallel algorithms for a special case of the $M$-RA, we derive algorithms that can split the computing task efficiently between many nodes. This way, spatial inference could be carried out for massive spatial datasets, using parallel computations at a number of nodes each dealing only with a subset of the dataset. If enough computing nodes are available, this ensures scalability of the $M$-RA even for datasets with many millions of observations.

This article is organized as follows. In Section \ref{sec:MRA}, we introduce the $M$-RA and discuss some of its properties. In Section \ref{sec:inference}, we present algorithms for parameter inference and spatial prediction, and describe the computational complexity of the $M$-RA. In Section \ref{sec:comparison}, we apply the $M$-RA to large simulated datasets and to a real-data example and compare the $M$-RA to the full-scale approximation. We conclude in Section \ref{sec:conclusions}. All proofs are given in Appendix \ref{app:proofs}.


\section{Multi-resolution approximation ($M$-RA) \label{sec:MRA}}

We begin this section by describing the true Gaussian process to be approximated (Section \ref{sec:general}). After some preliminaries (Section \ref{sec:knots}), we introduce the multi-resolution approximation ($M$-RA) (Section \ref{sec:mradef}) and discuss its properties (Sections \ref{sec:properties} and \ref{sec:choices}).

\subsection{The true Gaussian process \label{sec:general}}

Let $\{y_0(\bs) \!: \bs \in \domain\}$, or $y_0(\cdot)$, be the true spatial field or process of interest on a continuous (non-gridded) domain $\domain \subset \mathbb{R}^d$, $d \in \mathbb{N}^+$. We assume that $y_0(\cdot) \sim \GP(0,C_0)$ is a zero-mean Gaussian process with covariance function $C_0$. We place no restrictions on $C_0$, other than assuming that it is a positive-definite function on $\domain$ that is known up to a vector of parameters, $\bftheta$. In real applications, $y_0(\cdot)$ will often not have mean zero, but estimating and subtracting the mean is usually not a computational problem. Once $y_0(\cdot)$ has been observed at a set of $n$ spatial locations, $\locs = \{\bs_1,\ldots,\bs_n\}$, the distribution of the data is given by
\begin{equation*}
\label{eq:datadist}
  \by_0(\locs) \colonequals \big(y_0(\bs_1),\ldots,y_0(\bs_n)\big)' \sim N_n\big(\bfzero,\bC_0(\locs,\locs)\big),
\end{equation*}
an $n$-variate Gaussian distribution with covariance matrix $\bC_0(\locs,\locs) = \big(C_0(\bs_i,\bs_j)\big)_{i,j=1,\ldots,n}$\,. 

The basic goal of spatial statistics is to make (likelihood-based) inference on the parameters $\bftheta$ and to obtain spatial predictions of $y_0(\cdot)$ at a set of locations $\locs^P$ (i.e., to obtain the posterior distribution of $\by_0(\locs^P)$). This requires multiple Cholesky decompositions of the data covariance matrix $\bC_0(\locs,\locs)$, which generally has $\mathcal{O}(n^3)$ time and $\mathcal{O}(n^2)$ memory complexity. This is computationally infeasible when $n=10^5$ or more. In addition, computations for inference are also very difficult to parallelize, as computations with a dense and large covariance matrix require substantial communication overhead. Thus, the computational challenges cannot be solved by brute-force use of high-performance computing systems, and approximations or simplifying assumptions are necessary.

\subsection{Domain partitioning and knots \label{sec:knots}}

To define the $M$-RA, we need a recursive partitioning of the spatial domain $\domain$, in which each of $J$ regions, $\domain_1,\ldots,\domain_J$, is again divided into $J$ smaller subregions, and so forth, up to level $M$:
\[
\domain_{j_1,\ldots,j_{m-1}} = \dot{\textstyle\bigcup}_{j_{m}=1,\ldots,J} \, \domain_{j_1,\ldots,j_{m}}, \quad j_1,\ldots,j_{m-1}=1,\ldots,J; \quad m=1,\ldots,M,
\]
For a generic Gaussian process $x(\cdot) \sim \GP(0,C)$, we define $[x(\cdot)]_{[m]}$ to be a ``block-independent'' version of $x(\cdot)$ between regions at resolution $m$; that is, $[x(\cdot)]_{[m]}\sim \GP(0,[C]_{[m]})$, where $[C]_{[m]}(\bs_1,\bs_2) = C(\bs_1,\bs_2)$ if $\bs_1,\bs_2$ are in the same region $\domain_\jm$ at resolution $m$, and $[C]_{[m]}(\bs_1,\bs_2) =0$ otherwise.

\begin{figure}
\begin{subfigure}{1\textwidth}
	\begin{subfigure}{.48\textwidth}
	\centering
	\includegraphics[width =1\linewidth]{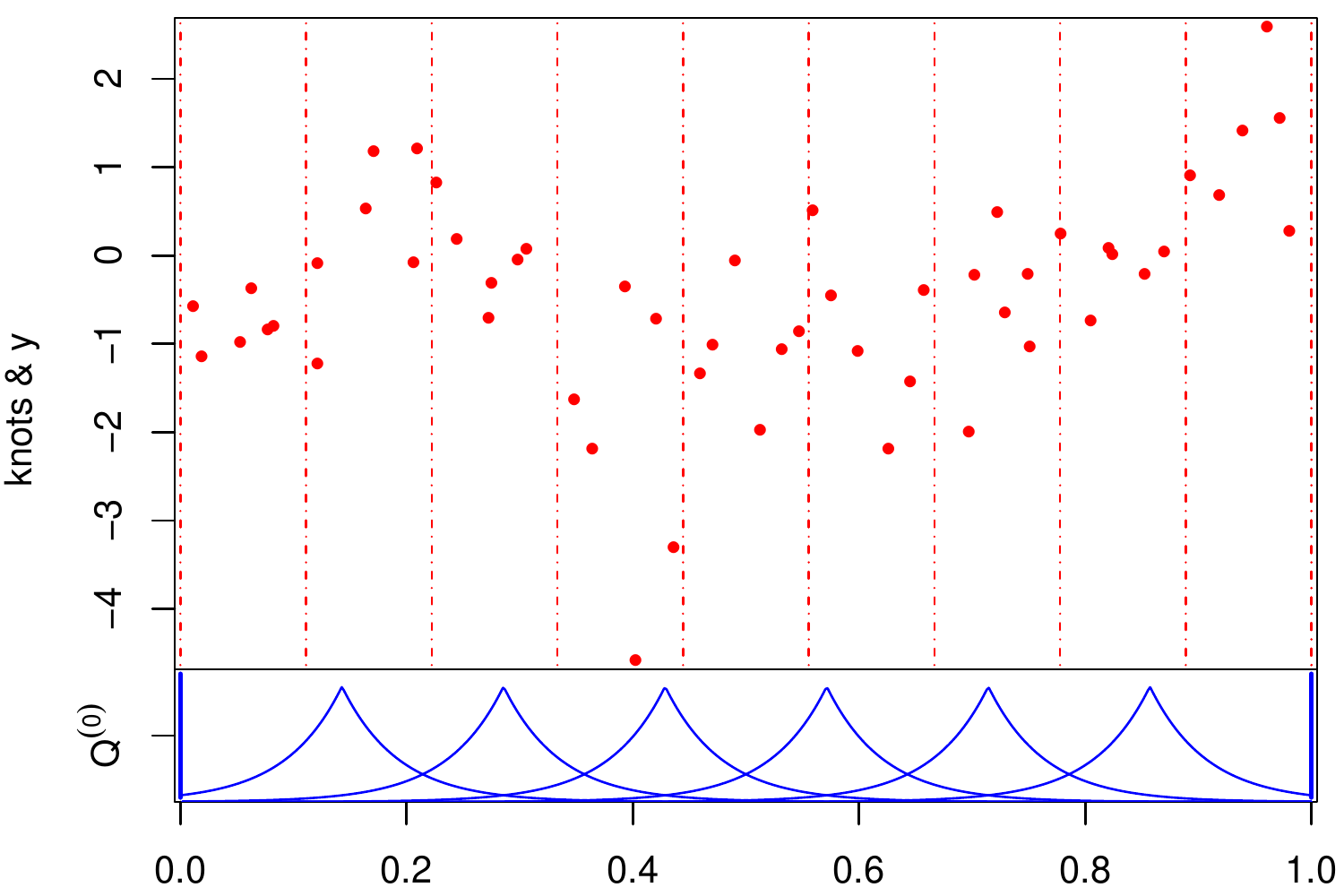}
	\caption{Simulated data (red dots), and $J=9$ partitions (red vertical lines) and $r=6$ basis functions (blue) for a 1-RA}
	\label{fig:toyfsadata}
	\end{subfigure}%
\hfill
	\begin{subfigure}{.48\textwidth}
	\centering
	\includegraphics[width =1\linewidth]{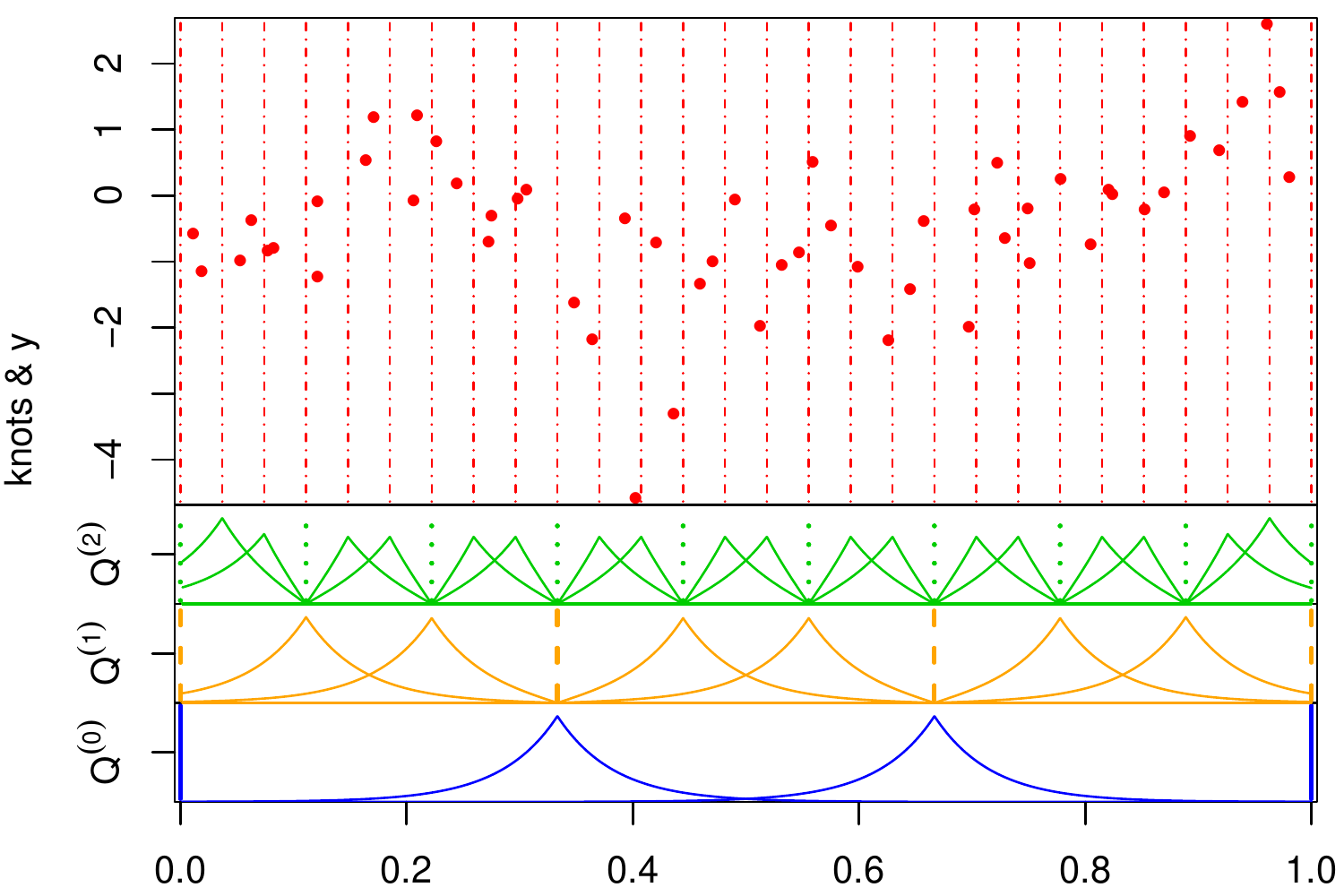}
	\caption{Data (red dots), and $J=3$ partitions (vertical lines) and $r=2$ basis functions per region for a 3-RA}
	\label{fig:toymradata}
	\end{subfigure}%
\end{subfigure}\\
\begin{subfigure}{1\textwidth}
	\begin{subfigure}{.48\textwidth}
	\centering
	\includegraphics[width =1\linewidth]{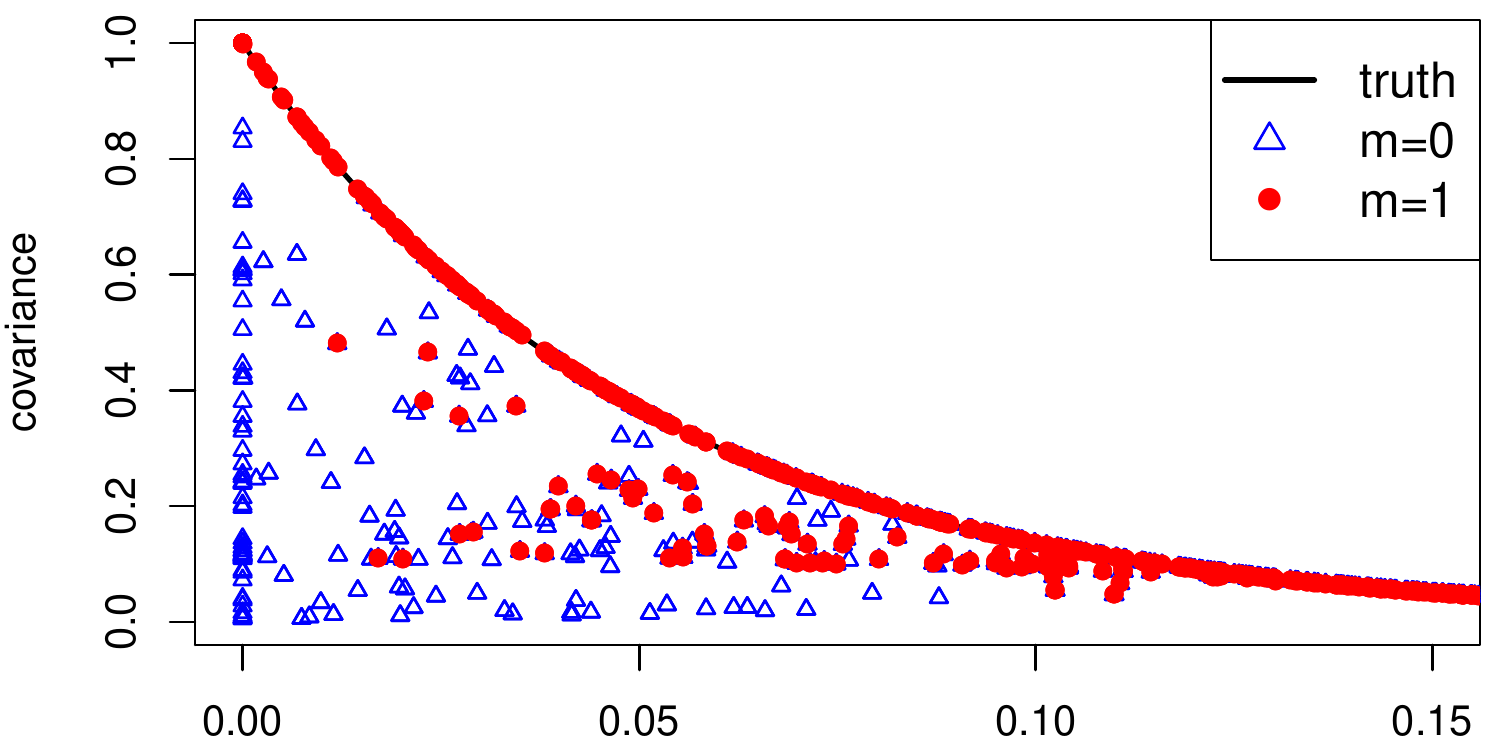}
	\caption{True covariance and 1-RA up to resolution $m$}
	\label{fig:FSAcov}
	\end{subfigure}%
\hfill
	\begin{subfigure}{.48\textwidth}
	\centering
	\includegraphics[width =1\linewidth]{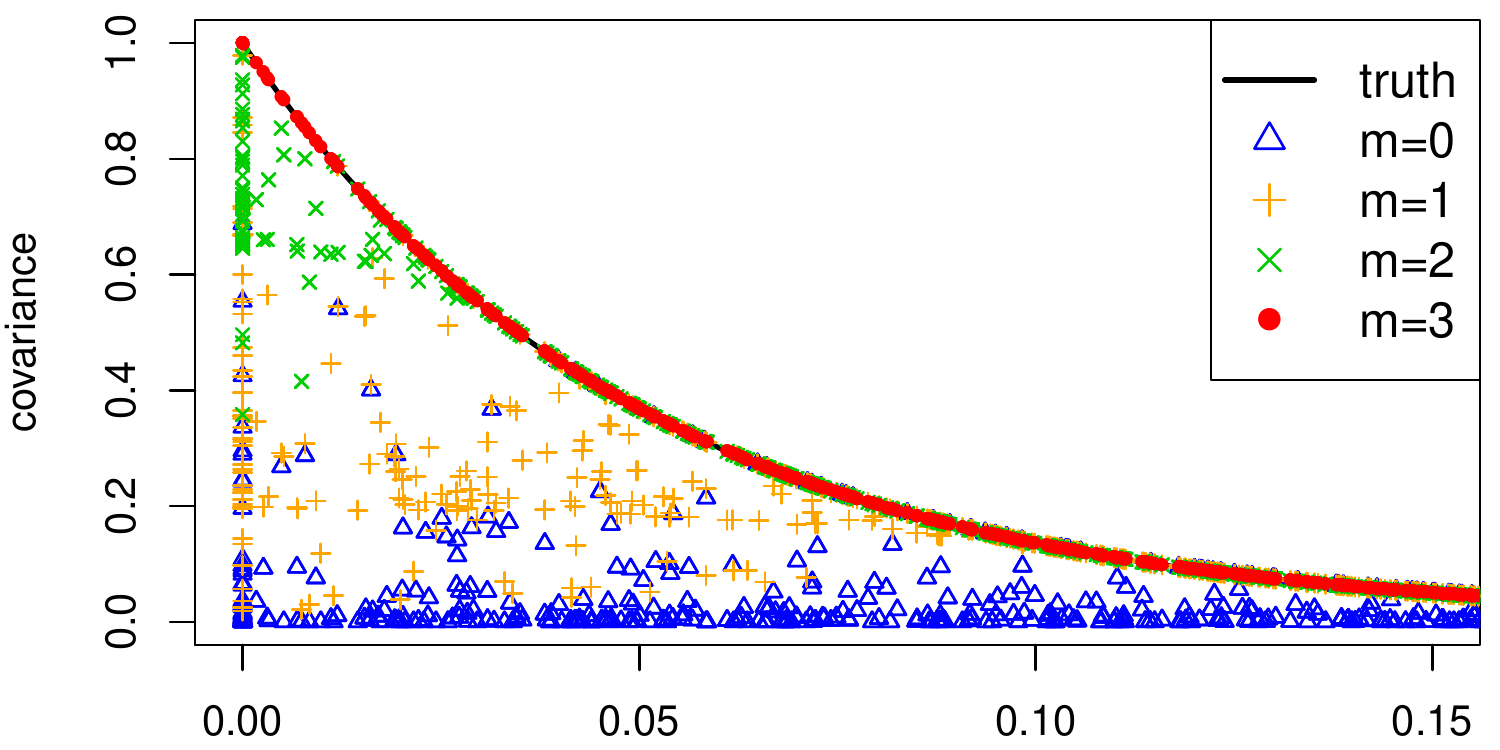}
	\caption{True covariance and $M$-RA up to resolution $m$}
	\label{fig:MRAcov}
	\end{subfigure}%
\end{subfigure}\\
\begin{subfigure}{1\textwidth}
\centering 
\includegraphics[width =.55\textwidth]{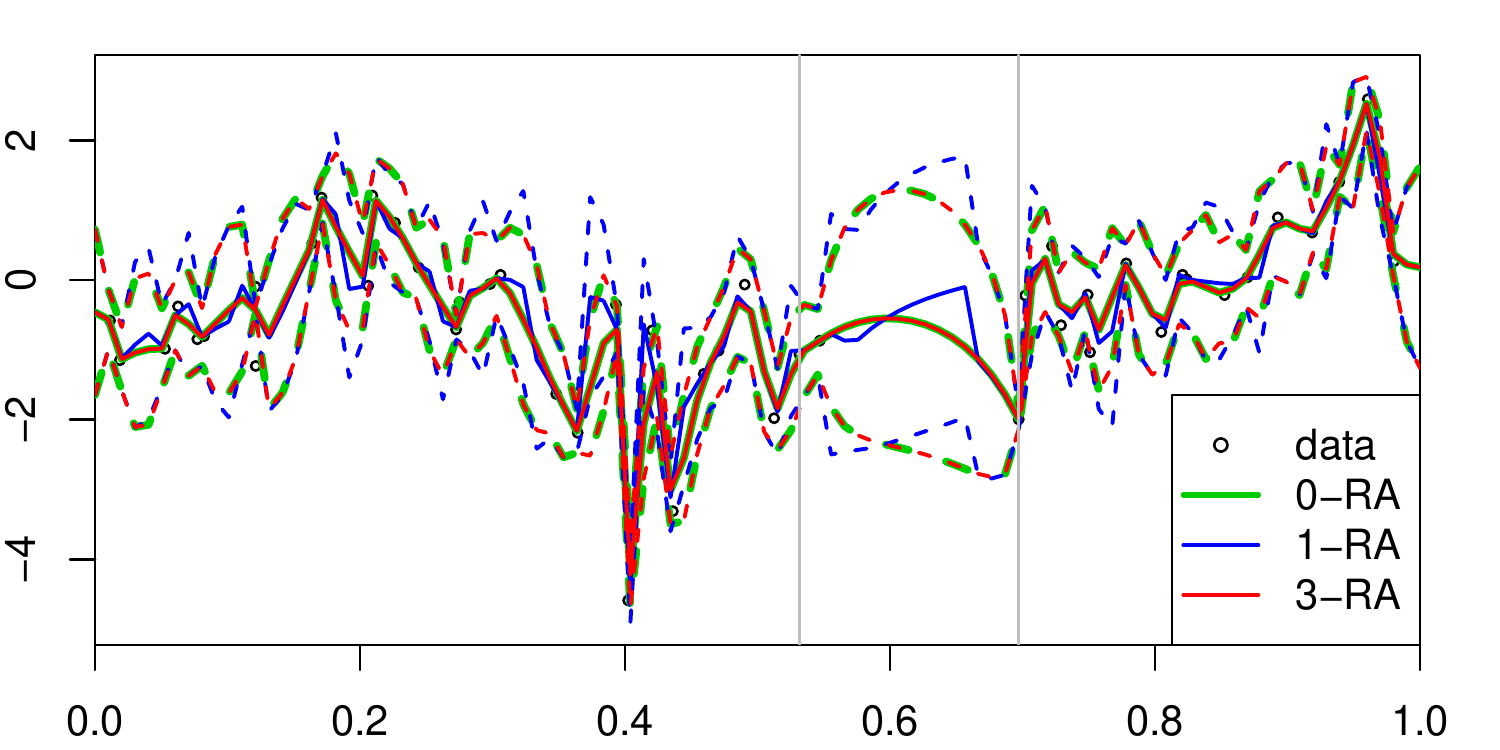}
\caption{Posterior predictive means (solid lines) and pointwise $95\%$ posterior prediction intervals (dashed lines) for $y(\cdot)$. The interval bounded by the gray vertical lines is ``unobserved.''}
\label{fig:predsillustration}
\end{subfigure}
\caption{Comparison of a full-scale approximation (1-RA) to a multi-resolution approximation with 3 resolutions (3-RA) with the same computational complexity ($Mr=6$ for both models) in a toy example of $n=54$ observations generated from an exponential covariance function on a one-dimensional spatial domain $\domain = [0,1]$. Panels (\protect\subref{fig:FSAcov}) and (\protect\subref{fig:MRAcov}) show the covariance approximations using only the basis functions up to resolution $m$, for $m=0,\ldots,M$. As $m$ increases, deviations between the true and approximated covariance occur only on smaller and smaller scales (distances). For $m=M$, the covariance and predictions of the 3-RA are exactly the same as the true covariance and the corresponding (optimal) predictions of the $0$-RA, whereas the covariance approximations and predictions using the 1-RA differ considerably from the truth. Note that for other covariance functions or in higher dimensions, the $M$-RA will not generally be exact (see Section \ref{sec:choices}).}
\label{fig:illustrationcomparison}
\end{figure}

We also need a multi-resolutional set of knots, such that $\knots_{j_1,\ldots,j_{m}}$ is a set of $r$ knots (with $r<<n$) that all lie in subregion $\domain_{j_1,\ldots,j_{m}}$. For ease of notation, we assume that the knots in each of the regions at resolution $M$ are given by the observation locations in that region: $\knots_{j_1,\ldots,j_{M}} = \locs_{j_1,\ldots,j_{M}}$. Further, we write $\knots^{(m)} = \{\knots_\jm: \jm = 1,\ldots,J\}$ for the set of all $rJ^m$ knots at resolution $m$.
For a one-dimensional toy example, the top row of Figure \ref{fig:illustrationcomparison} shows partitions and knots for $M$-RAs with $M=1$ and $M=3$. The knots are the locations at which the basis functions attain their maximum.

Note that we have assumed the same number ($J$) of subregions in each partition and the same number ($r$) of knots in each subregion, but this is only for notational convenience and not necessary for the inference described later. Hereafter, we will assume the domain partitioning and knots to be fixed and known. Some further discussion of their choice is given in Section \ref{sec:choices}.

\subsection{Definition of the multi-resolution approximation ($M$-RA) \label{sec:mradef}}

Recall the true Gaussian process $y_0(\cdot) \sim \GP(0,C_0)$ from Section \ref{sec:general}. We assume temporarily that the covariance function $C_0$ is fully known (parameter inference is discussed in Section \ref{sec:parameter}). The $M$-RA process approximates $y_0(\cdot)$ and its covariance $C_0$ iteratively at resolutions $m=0,\ldots,M$, based on the knots and partitions specified in Section \ref{sec:knots}. At each resolution $m$, it approximates the ``remainder'' term --- the difference between $y_0(\cdot)$ and the approximations at lower resolutions --- using the predictive process \citep{Banerjee2008}, independently between regions $\domain_\jm$. As illustrated in Panel (\subref{fig:MRAcov}) of Figure \ref{fig:illustrationcomparison}, low $M$-RA resolutions captures variability at low frequencies (i.e., at large distances), resulting in remainder terms that exhibit variability on smaller and smaller scales as $m$ increases, and so approximating the remainder independently between finer and finer partitions causes little approximation error. 

More specifically, we begin with a predictive-process approximation of $y_0(\cdot)$,
$  \pp_0(\bs) \colonequals E\big(y_0(\bs) | \by_0(\knots^{(0)}) \big), \,\bs \in \domain$,
and we approximate the remainder process by assuming it to be independent between regions $\domain_1,\ldots,\domain_J$ at resolution 1: $\delta_1(\cdot) = [y_0(\cdot) - \pp_0(\cdot)]_{[1]}$ \citep{Sang2011a}. We then again approximate this remainder process as the sum of its predictive process, $\pp_1(\bs) = E\big(\delta_1(\bs) | \bfdelta_1(\knots^{(1)}) \big), \,\bs \in \domain$, plus the approximate remainder $\delta_2(\cdot) = [\delta_1(\cdot) - \pp_1(\cdot)]_{[2]}$, and so forth, up to level $M$. This leads to the following expression for the $M$-RA:
\begin{equation}
\label{eq:mra1}
  y_M(\cdot) = \pp_0(\cdot) + \pp_1(\cdot) + \ldots + \pp_{M-1}(\cdot) + \delta_M(\cdot),
\end{equation}
where $\pp_m(\bs) = E\big(\delta_m(\bs) | \bfdelta_m(\knots^{(m)}) \big), \,\bs \in \domain$, and $\delta_m(\cdot) = [\delta_{m-1}(\cdot) - \pp_{m-1}(\cdot)]_{[m]} \sim \GP(0,v_m)$.

An alternative expression for the $M$-RA in \eqref{eq:mra1} can be obtained by noting that, for $m=0,\ldots,M-1$, we can write each predictive process for $\bs \in \domain_\jm$ as a linear combination of basis functions \citep[cf.,][]{Katzfuss2012}, $\pp_m(\bs) = \bb_\jm(\bs)'\bfeta_\jm$ with $\bfeta_\jm \sim N_r(\bfzero,\bK_\jm)$, and
\begin{equation}
\label{eq:msadetails}
\begin{split}
 \bb_\jm(\bs) & \colonequals v_m(\bs,\knots_\jm), \quad \bs \in \domain_\jm\\  
 \bK_\jm^{-1} & \colonequals v_m(\knots_\jm,\knots_\jm)\\
v_{{m+1}}(\bs_1,\bs_2) & \colonequals v_{m}(\bs_1,\bs_2) - \bb_\jm(\bs_1)'\bK_\jm \bb_\jm(\bs_2), \quad \bs_1,\bs_2 \in \domain_{j_1,\ldots,j_{m}},
\end{split}
\end{equation}
where $v_{{m+1}}(\bs_1,\bs_2)=0$ if $\bs_1$ and $\bs_2$ are in different regions at the $m$th resolution, and we set $v_{0} = C_0$. Therefore, the $M$-RA can also be written as a linear combination of basis functions at $M$ resolutions $0,\ldots,M-1$, plus a remainder term at resolution $M$:
\begin{equation}
\label{eq:mradef}
  y_M(\bs) = \bb(\bs)'\bfeta + \bb_{j_1}(\bs)'\bfeta_{j_1} + \, \ldots \, + \bb_{j_1,\ldots,j_{M-1}}(\bs)'\bfeta_{j_1,\ldots,j_{M-1}} + \delta_M(\bs), \quad \bs \in \domain_{j_1,\ldots,j_{M}},
\end{equation}
where the weight vectors are independent of each other and of $\delta_M(\cdot) \sim \GP(0,v_M)$. Panels (\subref{fig:toyfsadata}) and (\subref{fig:FSAcov}) of Figure \ref{fig:illustrationcomparison} show the basis functions in the toy example. Once we have observed data at locations $\locs$, we can also write the remainder $\delta_M(\cdot)$ in \eqref{eq:mradef} as a linear combination of basis functions, $\delta_M(\bs) = \bb_\jM(\bs)'\bfeta_\jM, \, \bs \in \domain_\jM$, as in \eqref{eq:msadetails}, where $\knots_\jM = \locs_\jM$.

\subsection{Properties of the $M$-RA \label{sec:properties}}

\subsubsection*{Many basis functions}

In contrast to so-called low-rank approaches, which rely on a small or moderate number of basis functions and for which capturing small-scale variation is challenging \citep[see, e.g.,][]{Stein2013a}, the total number of basis functions for the $M$-RA with $M>0$ is larger than the number of measurements: $r_\textnormal{total}= r \sum_{m=0}^M J^M = rJ^M + r\sum_{m=0}^{M-1} J^m = n + r\frac{J^M-1}{J-1} > n$. This allows the $M$-RA to capture variation at all spatial scales, including very small scales.

\subsubsection*{Orthogonal decomposition}

Because the predictive process is a conditional expectation, which is a projection operator, the predictive process $\pp_m(\cdot)$ is independent of the remainder $\delta_m(\cdot) - \pp_m(\cdot)$, for all $m=0,\ldots,M-1$. Hence, we define the $M$-RA in \eqref{eq:mra1} as a sum of orthogonal components. In the form \eqref{eq:mradef}, the $M$-RA is a weighted sum of spatial basis functions, for which the weights $\bfeta_\jm$ are block-orthogonal in probability space, but two sets of basis functions $\bb_{j_1,\ldots,j_{m_1}}$ and $\bb_{i_1,\ldots,i_{m_2}}$ are only block-orthogonal in physical space if $\domain_{j_1,\ldots,j_{m_1}} \cap \domain_{i_1,\ldots,i_{m_2}} = \emptyset$.

\subsubsection*{Valid Gaussian process}

\begin{prop}
\label{prop:nnd}
The $M$-RA is a valid Gaussian process with a nonnegative definite covariance function, $C_M$.
\end{prop}

\subsubsection*{``Optimal'' basis functions}

At every resolution $m=0,\ldots,M-1$ and within every region $\domain_\jm$, the goal of the $M$-RA in \eqref{eq:mra1} is to approximate the remainder process $\delta_m(\cdot)$ as closely as possible, where
\begin{equation}
\label{eq:remainder}
\textstyle\delta_m(\cdot) = [\delta_{m-1}(\cdot) - \pp_{m-1}(\cdot)]_{[m]} = [y_{0}(\cdot) - \sum_{l=0}^{m-1} \pp_{l}(\cdot)]_{[m]}.
\end{equation}
Hence, in each region, $\delta_m(\cdot)$ in \eqref{eq:mra1} is the difference between the true process $y_0(\cdot)$ and the ``previous'' terms at lower resolutions, $\sum_{l=0}^{m-1} \pp_{l}(\cdot)$. We choose $\pp_m(\cdot)$ to be the predictive-process approximation of $\delta_m(\cdot)$. As this is a conditional expectation, $\pp_m(\cdot)$ is the function of $\bfdelta_m(\knots^{(m)})$ that minimizes the expected squared difference to $\delta_m(\bs)$, conditional on $\bfdelta_m(\knots^{(m)})$ \citep{Banerjee2008}.
Further, $\pp_m(\cdot)$ can be viewed as an approximation of the optimal rank-$r$ representation of $\delta_m(\cdot)$ within each region $\domain_\jm$, in that $\pp_m(\cdot)$ is the Nystr\"om approximation of the first $r$ terms in the Karhunen--Lo\`eve expansion of $\delta_m(\cdot)$ \citep{Sang2012}. This is further evidence that at each resolution the predictive process captures variability at the low frequencies, leaving mostly higher-frequency variability to be captured at higher resolutions within smaller subregions. For increasing $r$, $\pp_m(\cdot)$ will be increasingly close to $\delta_m(\cdot)$. In fact, if $\knots_\jm$ is equal to $\locs_\jm$, the set of observed locations in $\domain_\jm$, then it is straightforward to show that $\bftau_m(\locs_\jm) = \bfdelta_m(\locs_\jm)$. In this sense, $\tau_m(\cdot)$ (and its basis-function representation) are an ``optimal'' approximation of $\delta_m(\cdot)$. 

\begin{figure}
\centering
	\begin{subfigure}{.3\textwidth}
	\centering 
	\includegraphics[width=1\linewidth]{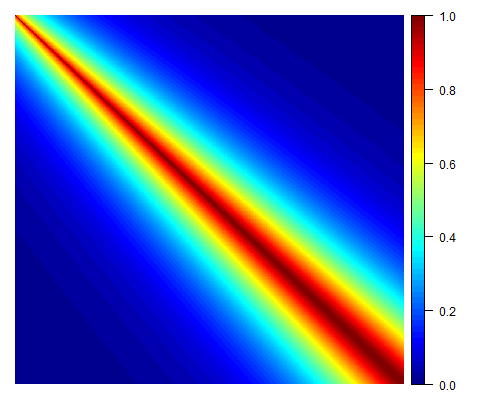}
	\caption{True covariance matrix}
	\label{fig:truecov1}
	\end{subfigure}%
\hfill
	\begin{subfigure}{.65\textwidth}
	\centering
	\includegraphics[width=.95\linewidth]{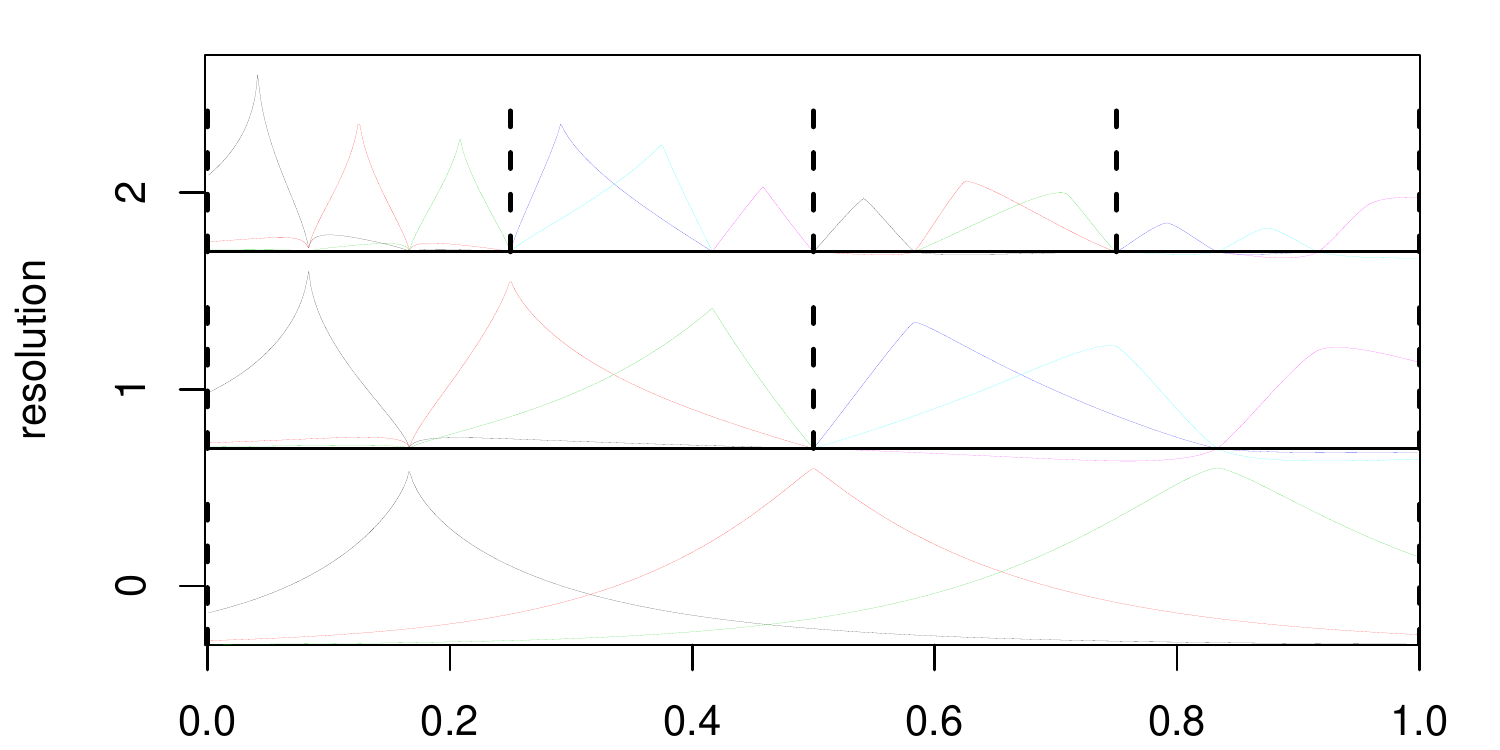}
	\caption{First 3 resolutions of basis functions and partitions (dashed lines)}
	\label{fig:basisfunctions}
	\end{subfigure}%
\caption{For a spatial process with a nonstationary Mat\'{e}rn covariance function with range 0.15 and spatially varying (increasing) smoothness $\nu(s) = 0.2 + 0.7s$ on a one-dimensional domain $\domain=[0,1]$, basis functions and partitions of a $M$-RA with $J=2$ and $r=3$. 
Note how the basis functions adapt to the increasing smoothness of the true covariance function and according to the placement of basis functions at lower resolutions.}
\label{fig:bfillustration}
\end{figure}

In contrast to many other multi-resolution methods for spatial data, the $M$-RA thus automatically provides a basis-function representation to approximate a given covariance function $C_0$ (based on a particular domain partitioning and set of knots), without any restrictions on $C_0$. This is illustrated in Figure \ref{fig:bfillustration}, which shows the basis functions of a 3-RA for a highly nonstationary covariance function $C_0$.

\subsubsection*{Quality of the covariance approximation}

At resolution $m$, the $M$-RA attempts to capture the covariance of the remainder $\delta_m(\cdot)$ between the partitions of each region $\domain_\jm$ by the predictive-process basis-functions component $\bb_\jm(\cdot)'\bfeta_\jm$. How close the covariance of the $M$-RA, $C_M(\bs_1,\bs_2)$, is to the true covariance, $C_0(\bs_1,\bs_2)$, depends on up to which resolution $\bs_1$ and $\bs_2$ lie in the same region. If $\bs_1$ and $\bs_2$ are in the same region at resolution $M$, then $C_M(\bs_1,\bs_2) = C_0(\bs_1,\bs_2)$. (To prove this, simply combine \eqref{eq:mra1} with \eqref{eq:remainder}.) This also implies that the variances of $y_0(\cdot)$ and $y_M(\cdot)$ are the same. If $\bs_1$ and $\bs_2$ are in the same region $\domain_\jm$ at resolution $m<M$, but not at resolution $m+1$, then $C_0(\bs_1,\bs_2)$ is only approximated by the basis functions up to resolution $m$: $C_M(\bs_1,\bs_2) = \sum_{l=0}^m \bb_{j_1,\ldots,j_l}(\bs_1)'\bK_{j_1,\ldots,j_l}\bb_{j_1,\ldots,j_l}(\bs_2)$.

\subsubsection*{Comparison to the full-scale approximation}

Important special cases of the $M$-RA are the original process $y_0$ for $M=0$, and the full-scale approximation \citep{Snelson2007,Sang2011a} for $M=1$. The full-scale approximation, or 1-RA, only has basis functions at one resolution with $r^\textnormal{F}$ knots $\knots^{\textnormal{F}}$, and a single level of domain partitioning, $\domain = \dot{\textstyle\bigcup}_{j=1,\ldots,J^{\textnormal{F}}} \domain^{\textnormal{F}}_{j}$. For massive datasets, the subregions $\domain^{\textnormal{F}}_{j}$ need to be very small to maintain computational feasibility. 

Comparing our $M$-RA (with $M>1$) and a full-scale approximation (1-RA) with $\knots^\textnormal{F} = \knots$ and $\{\domain_j^\textnormal{F}: j=1,\ldots,J^\textnormal{F}\} = \{ \domain_\jM: \jM =1,\ldots,J\}$, the covariance approximation for the two models is the same for pairs of locations in the same finest subregion $\domain_j^\textnormal{F}$ and for pairs in different subregions $\domain_i$ and $\domain_j$ with $i \neq j$ at the (coarsest) resolution 1. For all other pairs of locations, the $M$-RA has extra basis functions to capture their dependence, and if $r$ is sufficiently large that $\tau_m(\cdot)$ captures the dependence of $\delta_m(\cdot)$ between subregions $\domain_\jmp$ well, the $M$-RA will provide a better approximation of $y_0(\cdot)$ than the 1-RA.

As described later in Section \ref{sec:complexity}, the $M$-RA with $r$ knots has the same computational complexity as the 1-RA with $Mr$ knots. As illustrated in Figure \ref{fig:illustrationcomparison}, the $M$-RA can result in considerably better approximations. Further comparisons are presented in Section \ref{sec:comparison}.

\subsection{More on the choice of knots and partitions \label{sec:choices}}

To achieve good approximations, we recommend choosing $M$ and $J$ as small and $r$ as large as the computational resources allow (see later in Table \ref{tab:comp1}), under the constraint that $ r J^M \geq n$.

If the observation locations are approximately uniformly distributed over the domain $\domain$, the partitions can simply be obtained by recursively splitting each region into $J$ subregions of equal area. If the observation locations are far from uniform, more complicated partitioning schemes might be necessary to achieve fast inference.

The remaining issue is the placement of the $r$ knots within each region. A simple solution is to use equidistant grids over each region $\domain_\jm$, but it can also be advantageous to place more knots close to the boundaries within each region. To see why, remember from \eqref{eq:remainder} and Section \ref{sec:mradef} that the goal within each region $\domain_\jm$ is to approximate $\delta_m(\cdot) \sim \GP(0,v_m)$. Consider the case of a region with $J=2$ subregions containing observed locations $\locs = \{\locs_1,\locs_2\}$ with $\locs_j = \{\locs_j^B,\locs_j^I\}$, where $\locs^B_j$ are the locations within a distance $c$ from the boundary and $\locs_j^I$ are the remaining locations in the interior of subregion $j$. Choosing the knots $\knots_\jm = \locs^B \colonequals \{\locs_1^B,\locs_2^B\}$, it can be shown that $\var(\bfdelta_m(\locs)) = \var(\bftau_m(\locs)) + \var(\bfdelta_m(\locs)|\bfdelta_m(\locs^B))$, the latter being a matrix with only one nonzero block, $\var(\bfdelta_m(\locs^I)|\bfdelta_m(\locs^B))$. The only part of this matrix that is ignored by the $M$-RA is $\cov(\bfdelta_m(\locs^I_1),\bfdelta_m(\locs^I_2)|\bfdelta_m(\locs^B))$, which should be very small if $c$ is large and/or the screening effect \citep[e.g.,][]{Stein2011b} holds for $v_m$.

An extreme case of this strategy is illustrated in Figure \ref{fig:illustrationcomparison}. For the exponential covariance function without nugget in one spatial dimension, the screening effect holds exactly, in that two observations are conditionally independent given a third observation that separates the two. Because the knots for a particular resolution in Figure \ref{fig:illustrationcomparison} are placed on the boundaries between partitions at the next higher resolution, the $M$-RA is exact in this case. For covariance functions without screening effect or in higher dimensions, the $M$-RA will generally not be exact.

While the (favorable) numerical results in Section \ref{sec:simstudy} are obtained with the simplest choice of equal-area partitions and equally spaced knots, it is possible to adopt more complicated strategies, such as choosing the knots and partitions based on clustering \citep[e.g.,][]{Snelson2007} or using reversible-jump Markov chain Monte Carlo \citep[e.g.,][]{Gramacy2008,Katzfuss2012}. Any potential boundary effects due to the choice of partitions can be alleviated by carrying out several $M$-RAs with different, shifted partitions and combining the results using Bayesian model averaging \citep[e.g.,][]{Hoeting1999}.


\section{Inference \label{sec:inference}}

In this section, we describe inference for the $M$-RA. For a particular value of the parameter vector $\bftheta$, the covariance function $C_0$, and hence the basis functions $\bb_{j_1,\ldots,j_{m}}$ and the covariance matrices $\bK_{j_1,\ldots,j_{m}}$ in \eqref{eq:mradef} are fixed. The prerequisite for inference is to calculate the quantities summarizing the prior distribution induced by the $M$-RA at the chosen knots and observed locations (Section \ref{sec:prior}). Then, the main task for inference is to obtain the posterior distribution of the unknown weight vectors $\etaset_{M-1}$ (Section \ref{sec:mainalgorithm}), where we define $\etaset_{m} \colonequals \{\bfeta_\jl \! : \; j_1,\ldots,j_l=1,\ldots,J; \, l=0,\ldots,m \}$ for all $m=0,\ldots,M-1$ to be the set of all basis-function weights at resolution $m$ and all lower resolutions (and we let $\etaset_{-1}=\emptyset $ be the empty set). Once this posterior distribution has been obtained, it can be used to evaluate the likelihood (Section \ref{sec:parameter}) and to obtain spatial predictions (Section \ref{sec:prediction}). By exploiting the block-sparse multi-resolution structure of the prior and posterior precision matrices of the weights, we obtain inference that has excellent time and memory complexity (Section \ref{sec:complexity}), can take full advantage of distributed-memory systems with a large number of nodes (Section \ref{sec:distributed}), and is thus scalable to massive spatial datasets.

\subsection{Calculating the prior quantities \label{sec:prior}}

Let $\locs_\jm$ be the observation locations that lie in region $\domain_\jm$, and define
\begin{equation}
\label{eq:matrices}
\begin{split}
  \bB_\jm^l & \colonequals \bb_\jl(\locs_\jm), \quad l=0,1,\ldots,m,\\
  \bfSigma_\jm & \colonequals \var(\by_M(\locs_\jm) | \etaset_{m-1}) = \bB_\jm^m\bK_\jm\bB_\jm^m{}' + \bV_\jm, \\
  \bV_\jm & \colonequals \var(\by_M(\locs_\jm) | \etaset_{m}) =\blockdiag\{\bfSigma_{\jm,1},\ldots,\bfSigma_{\jm,J}\},
\end{split}
\end{equation}
for $m=0,1,\ldots,M-1$, and $\bfSigma_\jM \colonequals v_M(\locs_\jM,\locs_\jM)$.

For inference, we explicitly need to calculate the matrices $\{\bK^{-1}_\jm\!\!: \jm = 1,\ldots, J; m =0,\ldots,M-1\}$, $\{\bB^l_\jM\!\!: \jM = 1,\ldots, J; l=0,\ldots,M-1 \}$, and $\{\bfSigma_\jM\!\!: \jM = 1,\ldots, J\}$. Defining $\bW_\jm^l \colonequals v_l(\knots_\jm,\knots_\jl)$, we can do so by calculating 
\begin{equation}
\label{eq:priorw}
\textstyle\bW_\jm^l = C_0(\knots_\jm,\knots_\jl) - \sum_{k=0}^{l-1} \bW_\jm^k \bK_\jk \bW_\jl^k{}'
\end{equation}
for $m = 0,\ldots,M$, $\jm = 1,\ldots,J$, and $l = 0,\ldots,m$. Then we have $\bK_\jm^{-1} = \bW_\jm^m$ for $m<M$, $\bB_\jM^l = \bW_\jM^l$ for $l<M$, and $\bfSigma_\jM = \bW_\jM^M$.

As an aside, other parameterizations of these matrices (and the quantities in \eqref{eq:mradef}) are also possible and will lead to similar inference algorithms as described later, as long as the weight vectors are a priori independent and the basis functions have the same limited support.

\subsection{The posterior distribution of the basis-function weights \label{sec:mainalgorithm}}

The definition of the $M$-RA in \eqref{eq:mradef}, together with the definitions in \eqref{eq:matrices}, imply that
\[
  \textstyle\by_M(\locs_\jm) | \etaset_m \sim N(\sum_{l=0}^{m}\bB_\jm^l \bfeta_\jl, \bV_\jm).
\]
Using the results in \citet[][Sect.~3]{Katzfuss2014}, it can be shown that the conditional posterior distributions of the weight vectors for all $m=0,\ldots,M-1$ are given by 
\begin{equation}
\label{eq:posteriorgeneral}
\bfeta_{j_1,\ldots,j_{m}} | \by_M(\locs),\etaset_{m-1} \stackrel{ind.}{\sim} N_r(\widetilde{\bfnu}_{j_1,\ldots,j_m},\widetilde{\bK}_{j_1,\ldots,j_m}), \quad j_1,\ldots,j_m=1,\ldots,J,
\end{equation}
with posterior precision and mean
\begin{equation}
\label{eq:posterior}
\begin{split}
  \widetilde{\bK}_\jm^{-1} & \textstyle = \bK_\jm^{-1} + \bB_\jm^m{}' \bV_\jm^{-1} \bB_\jm^m = \bK_\jm^{-1} + \bA^{m,m}_\jm, \\
  \widetilde{\bfnu}_\jm & \textstyle = \widetilde{\bK}_\jm\big(\bB_\jm^m{}' \bV_\jm^{-1}(\by_M(\locs_\jm)-\sum_{l=0}^{m-1}\bB_\jm^l \bfeta_\jl)\big)\\
          & \textstyle = \widetilde{\bK}_\jm \bfomega_\jm^m - \sum_{l=0}^{m-1} \bK_\jm \bA^{m,l}_\jm \bfeta_\jl
\end{split}
\end{equation}
respectively, where
\begin{equation}
\label{eq:mainquantities}
\begin{split}
\textstyle\bA^{k,l}_\jm & \colonequals \textstyle\bB_\jm^k{}' \bV_\jm^{-1} \bB_\jm^l = \textstyle\sum_{j_{m+1}=1}^J \widetilde{\bA}_\jmp^{k,l},\\
\textstyle\bfomega^k_\jm & \colonequals \textstyle\bB_\jm^k{}' \bV_\jm^{-1}\by_M(\locs_\jm)= \textstyle\sum_{j_{m+1}=1}^J \widetilde{\bfomega}_\jmp^{k},
\end{split}
\quad k\geq l = 0,\ldots,m,
\end{equation}
can be obtained recursively for $m=M-1,M-2,\ldots,0$ using
\begin{equation}
\label{eq:helpquantities}
\begin{split}
\widetilde{\bA}_\jm^{k,l} & \colonequals \bB_\jm^k{}' \bfSigma_\jm^{-1} \bB_\jm^l = \bA^{k,l}_\jm - \bA^{k,m}_\jm \widetilde{\bK}_\jm \bA^{m,l}_\jm,\\
\widetilde{\bfomega}^k_\jm & \colonequals \bB_\jm^k{}' \bfSigma_\jm^{-1} \by(\locs_\jm) = \bfomega^{k}_\jm - \bA^{k,m}_\jm \widetilde{\bK}_\jm \bfomega^{m}_\jm.
\end{split}
\end{equation}
In practice, the quantities in \eqref{eq:helpquantities} are calculated directly from the definition (first equality) for $m=M$, and using the recursive expression (second equality) for $m=M-1,\ldots,0$. The proof of results \eqref{eq:mainquantities}--\eqref{eq:helpquantities} is straightforward using the Sherman-Morrison-Woodbury formula \citep{Sherman1950,Woodbury1950,Henderson1981}:
\begin{equation}
\label{eq:smw}
\bfSigma_\jm^{-1} = \bV_\jm^{-1} - \bV_\jm^{-1} \bB_\jm^m \widetilde{\bK}_\jm \bB_\jm^m{}'\bV_\jm^{-1}.
\end{equation}

\subsection{Parameter inference \label{sec:parameter}}

Inference for the $M$-RA is based on $L_M(\bftheta)$, the likelihood of the observations $\by_M(\locs) \sim N_n(\bfzero,\bfSigma)$, where $\bfSigma = \bfSigma(\bftheta)$ is the $M$-RA covariance matrix given in \eqref{eq:matrices} with $m=0$ based on parameter vector $\bftheta$:
\begin{equation*}
\label{eq:approxlikelihood}
 - 2 \log L_M(\bftheta) = \log |\bfSigma| + \by_M(\locs)'\,\bfSigma^{-1}\,\by_M(\locs).
\end{equation*}
This likelihood can be calculated using the quantities in Section \ref{sec:mainalgorithm}. We have $- 2 \log L_M(\bftheta) = d+ u$, with 
\begin{equation*}
\begin{split}
d_\jm & \colonequals \log | \bfSigma_\jm |,\\
u_\jm & \colonequals \by_M(\locs_\jm)'\bfSigma_\jm^{-1}\by_M(\locs_\jm),
\end{split}
\qquad m = 0,1,\ldots,M.
\end{equation*}
For $m=M$, these quantities are calculated using the definition. For $m=M-1,\ldots,0$, the recursive expressions
\begin{align*}
\textstyle d_\jm & = \textstyle\log |\widetilde{\bK}_\jm^{-1}| - \log |\bK_\jm^{-1}| +\sum_{j_{m+1}=1}^J d_\jmp, \\
\textstyle u_\jm & = \textstyle - \bfomega^m_\jm{}'\widetilde{\bK}_\jm \bfomega^m_\jm + \sum_{j_{m+1}=1}^J u_\jmp,
\end{align*}
can be derived using a matrix determinant lemma \citep[e.g.,][Thm.~18.1.1]{Harville1997} and the Sherman-Morrison-Woodbury formula in \eqref{eq:smw}.

In summary, the $M$-RA log-likelihood can be written as a sum of log-determinants and quadratic forms involving only $r \times r$ matrices. This result enables fast and scalable evaluation of the likelihood, which in turn allows for a wide array of likelihood-based inference techniques for an unknown parameter vector $\bftheta$, such as maximum-likelihood estimation, Markov chain Monte Carlo, or particle filtering in spatio-temporal contexts.

\subsection{Spatial prediction \label{sec:prediction}}

Spatial prediction can be carried out separately, after parameter inference is completed. In a frequentist context, prediction only has to be carried out once, for the final parameter estimates. In a Bayesian framework, parameter inference can be carried out only for the thinned MCMC chain, or for particles with considerable weight in the case of a particle sampler.

Implicitly conditioning on a particular value of the parameter vector $\bftheta$, spatial prediction amounts to finding the posterior predictive distribution, $\by_M(\locs^P) | \by_M(\locs)$, at a set of prediction locations $\locs^P$. We denote by $\locs_\jM^P$ the prediction locations in region $\domain_\jM$. As a first step, we need to calculate prior prediction quantities similar to \eqref{eq:priorw},
\[
\bU_\jM^l \colonequals \textstyle v_l(\locs_\jM^P,\knots_\jl) = C_0(\locs_\jM^P,\knots_\jl) - \sum_{k=0}^{l-1} \bU_\jM^k \bK_\jk \bW_\jl^k{}',
\]
for $l=0,\ldots,M$, and then we set $\bL_\jM^M \colonequals v_M(\locs^P_\jM,\locs_\jM) = \bU_\jM^M$ and 
\begin{align*}
\bV_\jM^P & \colonequals \textstyle v_M(\locs^P_\jM,\locs^P_\jM) = C_0(\locs_\jM^P,\locs_\jM^P) - \sum_{k=0}^{l-1} \bU_\jM^k \bK_\jk \bU_\jM^k{}'.
\end{align*}
Spatial predictions can then be obtained using the following proposition.

\begin{prop}
\label{prop:prediction}
The posterior predictive distribution can be written as
\begin{equation}
\label{eq:prediction}
  \by_M(\locs^P_\jM) | \by_M(\locs) \, = \, \textstyle \sum_{m=0}^{M-1} \widetilde{\bB}^{m+1,m}_\jM \widetilde{\bfeta}_\jm + \widetilde{\bfdelta}_\jM,
\end{equation}
where
\begin{align*}
\widetilde{\bfeta}_\jm & \stackrel{ind.}{\sim} N_r( \widetilde{\bK}_\jm \bfomega^m_\jm, \widetilde{\bK}_\jm ),\\
\widetilde{\bfdelta}_\jM & \stackrel{ind.}{\sim} N(\bL_{\jM}^M\bfSigma^{-1}_\jM \by_M(\locs_\jM),\bV^P_\jM - \bL_{\jM}^M\bfSigma^{-1}_\jM\bL_{\jM}^M{}'),
\end{align*}
and the ``posterior basis-function matrices'' are given by 
\begin{equation}
\label{eq:bfrecursion}
\widetilde{\bB}^{l,k}_\jM\colonequals \bb_\jk(\locs^P_\jM) - \bL_\jM^l \bfSigma^{-1}_\jl \bB^k_\jl = \widetilde{\bB}^{l+1,k}_\jM - \widetilde{\bB}^{l+1,l}_\jM \widetilde{\bK}_\jl \bA^{l,k}_\jl,
\end{equation}
for $k=0,1,\ldots,l-1$. 
\end{prop}

Hence, the posterior predictive distribution in \eqref{eq:prediction} has the same form as the (prior) $M$-RA process in \eqref{eq:mradef}. This allows calculation and storage of the \emph{joint posterior predictive distribution}. Often, interest is in summaries of this joint posterior predictive distribution, such as the marginal posterior predictive distributions at each prediction location. In practice, the posterior basis-function matrices in \eqref{eq:bfrecursion} are calculated directly from the definition (first equation) for $l=M$, and using the recursive relation (second equation) for $l=M-1,\ldots,0$.

\subsection{Distributed computing \label{sec:distributed}}

\begin{figure}
\centering
\includegraphics[width =.59\textwidth]{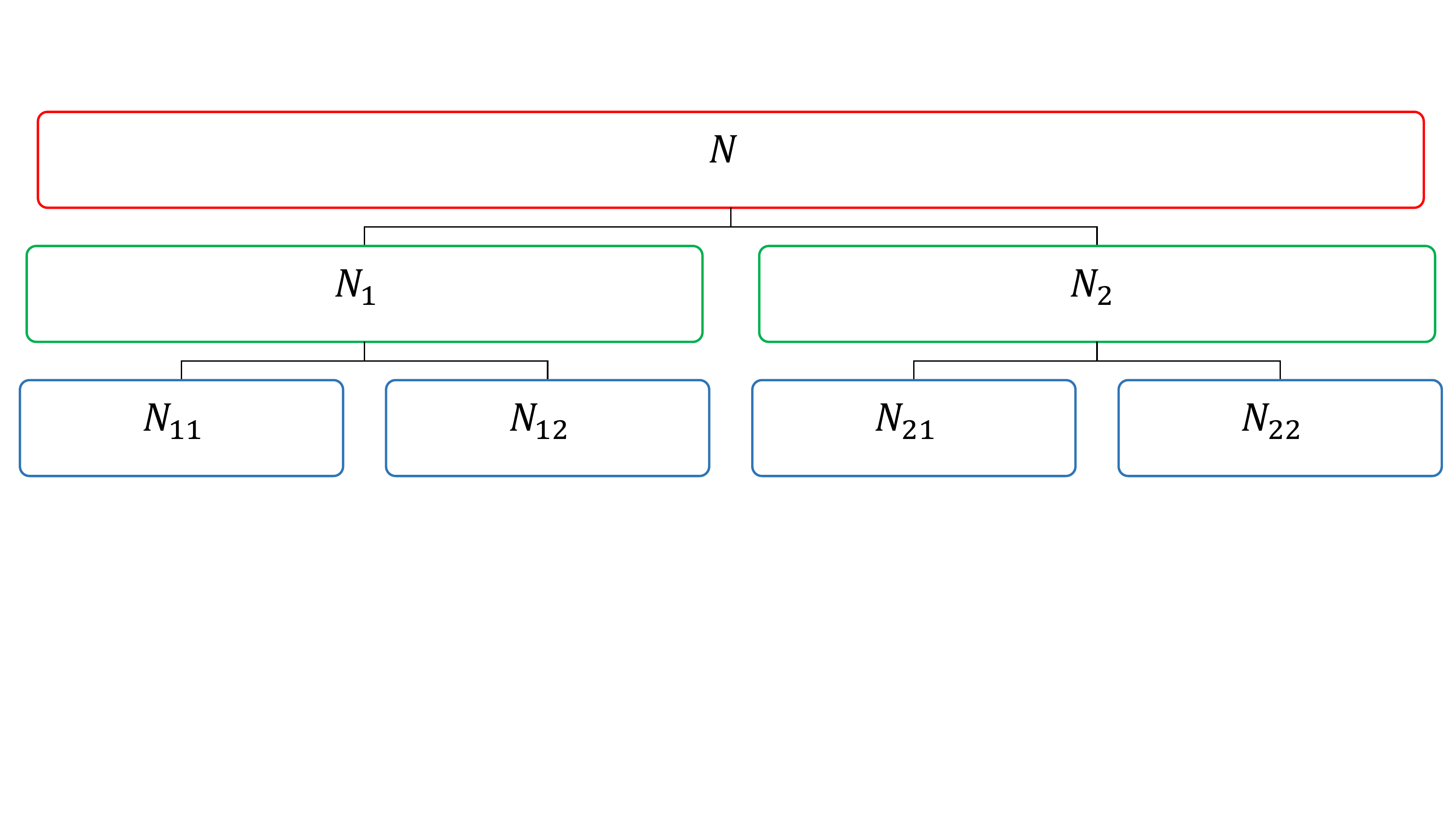}\\
\vspace{-23mm}
\includegraphics[width =.72\textwidth]{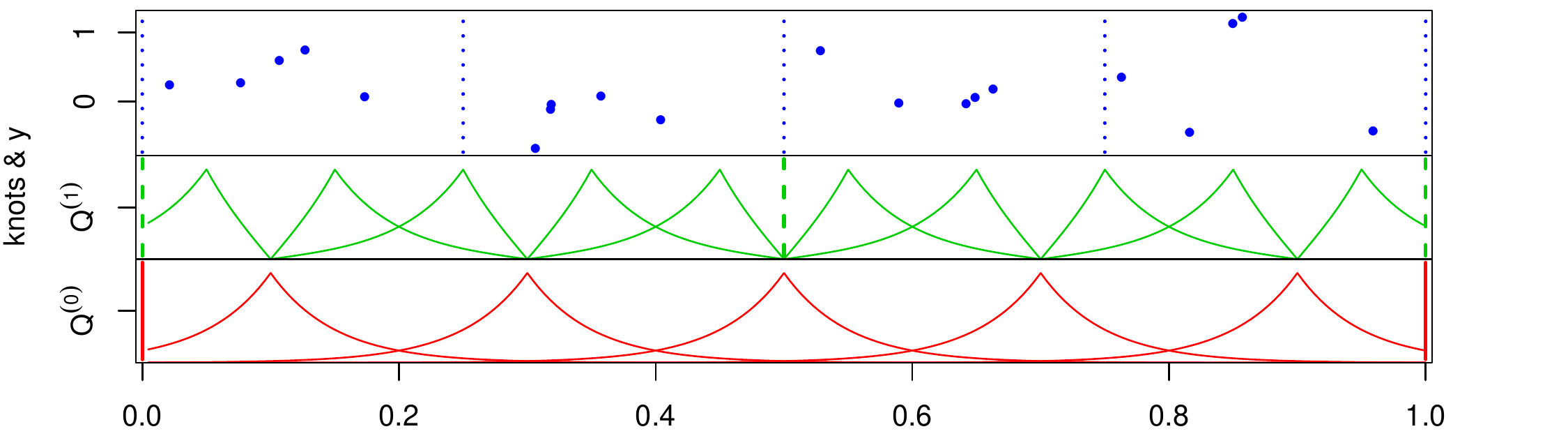}
\caption{Illustration of the computational setup for distributed inference in the $2$-RA with $J=2$ and $r=5$ for a toy example in a one-dimensional spatial domain. As indicated by matching colors, node $\node_{j_1,\ldots,j_m}$ works with the knots $\knots_{j_1,\ldots,j_m}$, $j_m=1,2$; $m=0,1,2$. Only communication between connected nodes is necessary. }
\label{fig:nodesIllustration}
\end{figure}


A major advantage of the $M$-RA is that it is well suited to modern computing environments, in that computations can be carried out in a distributed fashion with little communication overhead at a large number of nodes, each only dealing with a small subset of the data.

More specifically, assume that we have nodes $\{\node_{j_1,\ldots,j_{m}}\!: j_1,\ldots,j_m=1,\ldots,J; \; m=0,1,\ldots,M \}$ in a tree-like structure, as illustrated in Figure \ref{fig:nodesIllustration}. Each node $\node_{j_1,\ldots,j_{m}}$ holds the $r$ knots or observation locations $\knots_{j_1,\ldots,j_{m}}$ located in subregion $\domain_{j_1,\ldots,j_{m}}$, and it only has to work with matrices of size $r \times r$ (implying excellent load balance).  The main computational effort for node $\node_{j_1,\ldots,j_{m}}$ is in computing the Cholesky decomposition of the $r \times r$ matrix $\widetilde{\bK}_{j_1,\ldots,j_{m}}$ and calculating the quantities in \eqref{eq:helpquantities}, the latter of which could be parallelized if the node has multiple cores. The communication to each node $\node_{j_1,\ldots,j_{m}}$ is $\mathcal{O}(Jm^2r^2)$, as it receives the matrices to calculate the quantities in \eqref{eq:mainquantities} from its children. The calculations at the nodes/subregions for each resolution can be carried out completely in parallel.

For spatial prediction at locations $\locs^P$, each (terminal) node $\node_\jM$ carries out parallel computations involving $\locs^P_\jM$, the prediction locations in region $\domain_\jM$, to obtain the ``posterior basis-function matrices'' in \eqref{eq:bfrecursion}.


\subsection{Computational complexity \label{sec:complexity}}

Remember that we assume here for simplicity that there is an equal number of $r=\frac{n}{J^M}$ knots or observation locations in each region $\domain_{j_1,\ldots,j_{m}}$, and we regard $J$ as a fixed (small) number. For each of the $\sum_{m=0}^M J^m < 2J^M$ regions, the main computational effort for inference is in obtaining the matrices $\widetilde{\bA}_\jm^{k,l}$ in \eqref{eq:helpquantities}, which requires one Cholesky decomposition of the $r \times r$ matrix $\widetilde{\bK}_{j_1,\ldots,j_{m}}$ and computing $\mathcal{O}(m^2)$ quadratic forms of size $r \times r$. Thus, $M$-RA inference has $\mathcal{O}(J^MM^2r^3)=\mathcal{O}(nr^2M^2)$ time complexity and $\mathcal{O}(nrM)$ memory complexity. 

When the $M$-RA is implemented in a distributed environment with a large number of nodes (as in Section \ref{sec:distributed} above), the overall time complexity is $\mathcal{O}(M^3r^3)$ and the memory complexity per node is $\mathcal{O}(Mr^2)$, assuming that communication (which is $\order(M^2r^2)$ per node) does not dominate computation time.

\begin{table}
\centering
\begin{subfigure}{.48\textwidth}
\centering  \small
\begin{tabular}{l|cc|cc}
 & \multicolumn{2}{c|}{Single Processor}                 & \multicolumn{2}{c}{Distributed}       \\
 & Time                             & Memory            & Time             & Memory            \\ \hline
0-RA & $n^3 $                 & $n^2 $ &   &   \\
1-RA & $n r^2$ & $n r$         & $r^3$            & $r^2$            \\
$M$-RA & $n (Mr)^2$ & $nMr$        & $(Mr)^3$        & $Mr^2$       
\end{tabular}
\caption{Increasing $r$ and $M$}
\label{tab:comp1}
\end{subfigure}%
\hfill
\begin{subfigure}{.48\textwidth}
\centering \small
\begin{tabular}{l|cccc}
     & \multicolumn{2}{c|}{Single Processor} & \multicolumn{2}{c}{Distributed} \\
     & Time  & \multicolumn{1}{c|}{Memory}  & Time           & Memory        \\ \hline
$M$-RA & $n \log^2 n$    & \multicolumn{1}{c|}{$n \log n$}     & $\log^3 n$        & $\log n$          
\end{tabular}
\vspace{3mm}
\caption{$M=\mathcal{O}(\log n)$}
\label{tab:comp2}
\end{subfigure}
\caption{Time and memory complexity of the $M$-RA and its special cases, regular kriging ($0$-RA) and full-scale approximation (1-RA), on a single computer and in the distributed setting of Section \ref{sec:distributed}}
\label{tab:complexity}
\end{table}

Thus, the $M$-RA with $r$ knots has the same computational complexity as the 1-RA with $Mr$ knots, but the $M$-RA can provide a much better approximation (see Figure \ref{fig:illustrationcomparison}). As is further explored in Section \ref{sec:simstudy} below, as $n$ is increasing, the performance of the 1-RA degrades unless $r$ is allowed to increase as some fraction of $n$, while for the $M$-RA we can keep $r$ fixed and instead let $M$ increase with $n$ as $M=\mathcal{O}(\log_J n)$. This is a natural assumption under increasing-domain asymptotics, for which an increase in the domain and data size by a factor of $J$ allows an additional split of the resulting domain (i.e., increasing $M$ by one) without degrading the approximation within the $J$ subregions at the first resolution. In this case, the time and memory complexity for the $M$-RA are $\mathcal{O}(n\log^2 n)$ and $\mathcal{O}(n\log n)$, respectively, in the non-distributed setting. In the distributed setting, the overall time complexity is then $\mathcal{O}(\log^3 n)$, and the memory and communication complexity per node are $\mathcal{O}(\log n)$ and $\mathcal{O}(\log^2 n)$, respectively.

The same complexities hold for prediction, as long as the number of prediction locations per terminal region is on the same order as the number of observed locations (i.e., $|\locs^P_\jM| = \mathcal{O}(r)$). In addition, the $M$-RA allows us to store the entire joint predictive distribution in $\mathcal{O}(Mr^2J^M) = \mathcal{O}(nMr)$ memory ($\mathcal{O}(Mr^2)$ per node in the distributed case). If $M=\mathcal{O}(\log n)$, this is $\mathcal{O}(n \log n)$ (or $\mathcal{O}(\log n)$ per node).

As summarized in Table \ref{tab:complexity}(\subref{tab:comp2}), if we let $M$ increase as $\log n$, the time and memory complexity for the $M$-RA are both quasilinear in $n$, and even polylogarithmic in distributed settings with many nodes.  Hence, the $M$-RA is highly scalable and can handle truly massive spatial datasets if enough computational nodes are available.


\section{Numerical comparisons and illustrations \label{sec:comparison}}

Using simulated and real data, we compared our proposed $M$-RA to the full-scale approximation (1-RA), which is a special case of the $M$-RA and a current state-of-the-art method for large spatial data. A non-distributed implementation of the methods in Julia (\url{http://julialang.org/}) version 0.3.7 was run on a 16-core machine (Intel Xeon 2.90GHz) with 64GB RAM. All Julia code, R code to produce the plots, and the data for Section \ref{sec:tpw} are available as supplementary material.

\subsection{Simulation study \label{sec:simstudy}}

We simulated five datasets, each roughly of size 2 million (specifically, $n_\textnormal{max}= $ 1,966,080) from a Gaussian process with mean zero and covariance function
\[
 C_0(s_1,s_2) = 0.95 \, \mathcal{M}_{1.5}(|s_1-s_2|/0.05) + 0.05 \, I(s_1 = s_2), \quad s_1, s_2 \in \domain,
\]
where $I(\cdot)$ is the indicator function and
\begin{equation}
\label{eq:mat15}
 \mathcal{M}_{1.5}(h) = \big(1 + h\sqrt{3}\,\big)\exp\big(-h\sqrt{3}\,\big), \quad h \in \mathbb{R}^+_0
\end{equation}
is a Mat\'{e}rn correlation function with smoothness parameter $1.5$, which is also used for the real-data example below in Section \ref{sec:tpw}. The data were simulated on an equidistant grid on a one-dimensional domain $\domain = [0,1]$, which permitted fast simulation using the Davies-Harte algorithm and evaluation of the exact likelihood using the Durbin-Levinson algorithm for comparison. These algorithms were implemented in Julia along the lines of the functions \texttt{DHsimulate} and \texttt{DLLoglikelihood} in the R package \texttt{ltsa} \citep{McLeod2007}.

From the ``full'' datasets of size $n_\textnormal{max}$, we created datasets of varying sample sizes $n$ roughly between 2,000 and 2 million. We considered the two types of asymptotics commonly used in spatial statistics: For fixed-domain (infill) asymptotics, we took equally spaced subsets of size $n$ from the full dataset on the entire domain $[0,1]$. For increasing-domain asymptotics, we always took the first $n$ observations from the entire set.

\begin{figure}
\begin{subfigure}{1\textwidth}
	\begin{subfigure}{.48\textwidth}
	\centering
	\includegraphics[trim = 0mm 3mm 0mm 0mm, clip, width =1\linewidth]{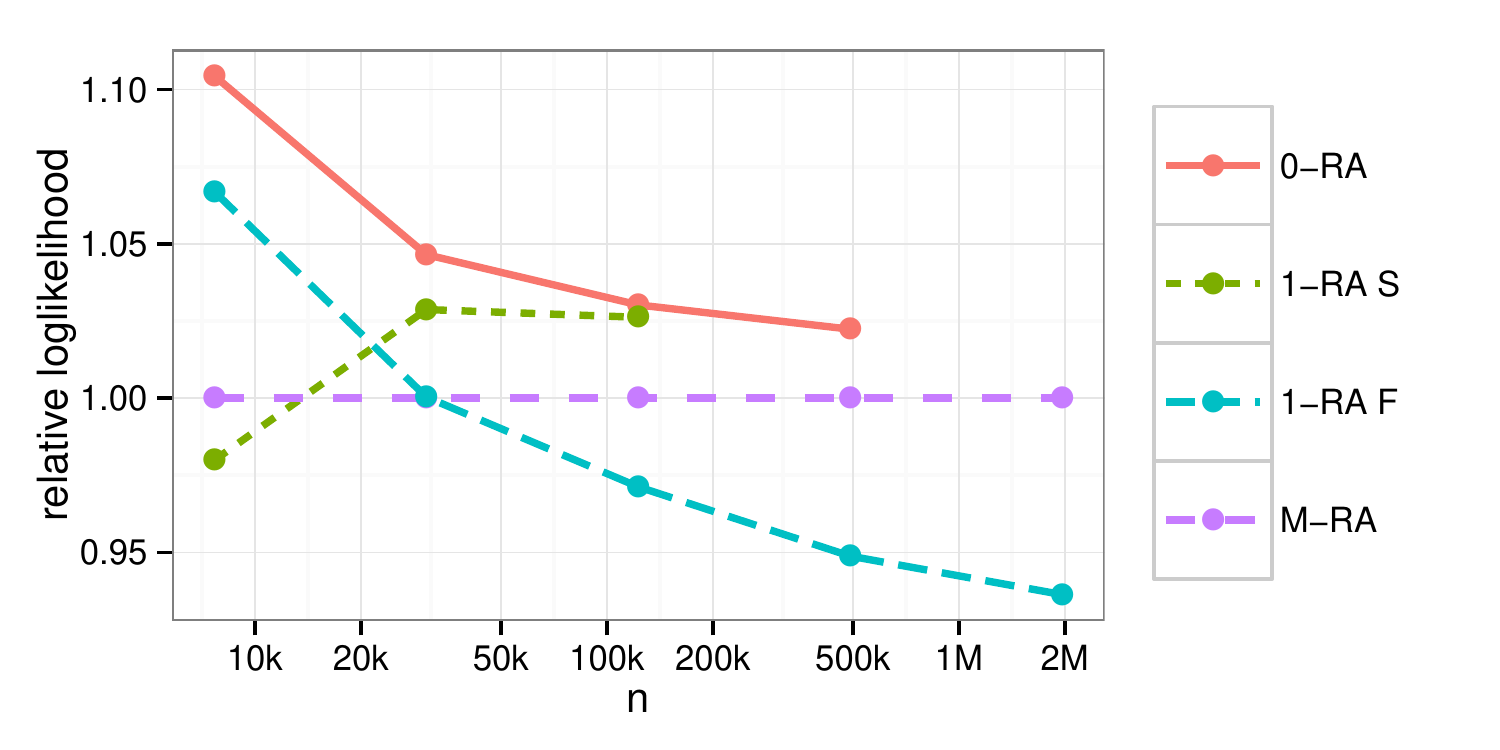}
	\caption{Loglikelihood under fixed domain}
	\label{fig:simfixed}
	\end{subfigure}
\hfill
	\begin{subfigure}{.48\textwidth}
	\centering
	\includegraphics[trim = 0mm 3mm 0mm 0mm, clip, width =1\linewidth]{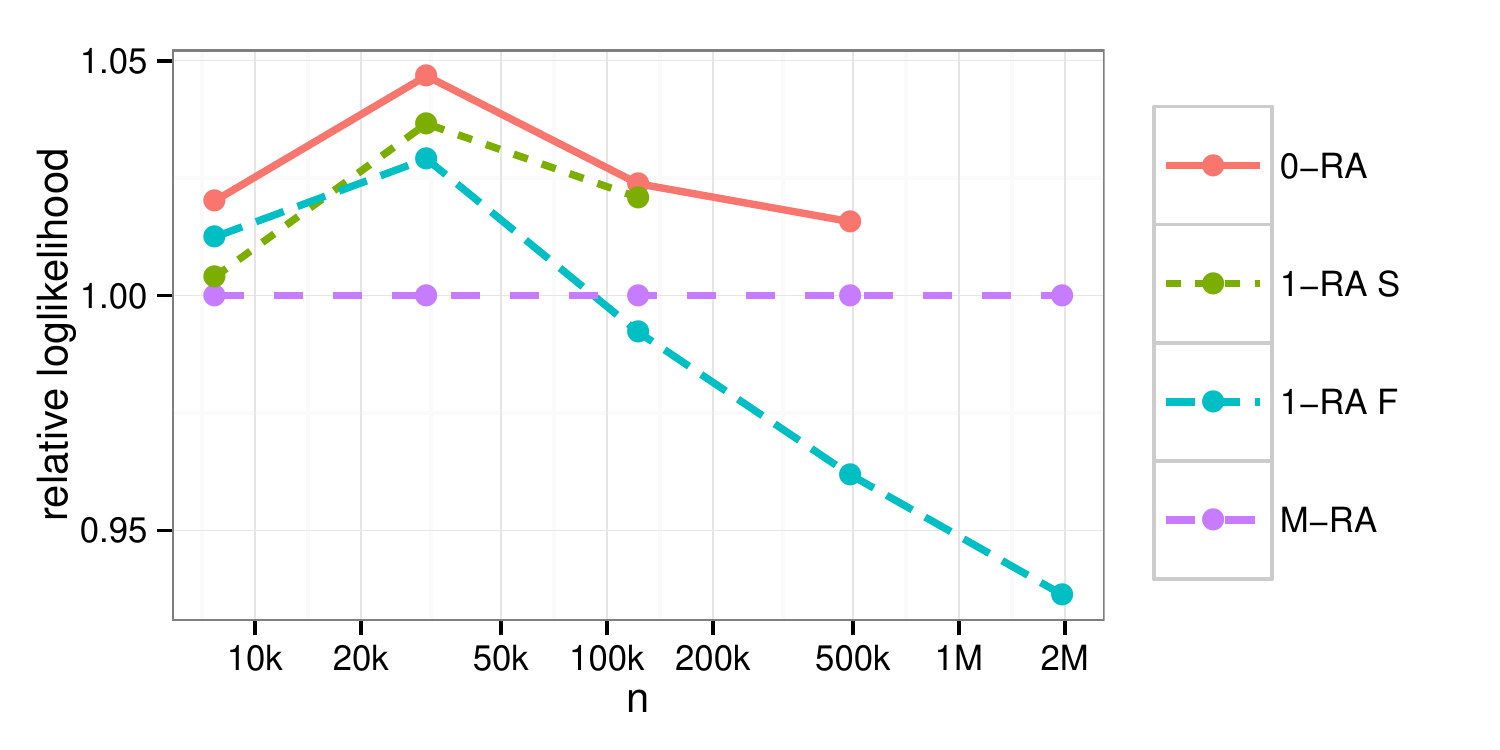}
	\caption{Loglikelihood under increasing domain}
	\label{fig:simincr}
	\end{subfigure}%
\end{subfigure}

\vspace{3mm}

\begin{subfigure}{1\textwidth}
	\begin{subfigure}{.48\textwidth}
	\centering
	\includegraphics[trim = 0mm 3mm 0mm 0mm, clip, width =1\linewidth]{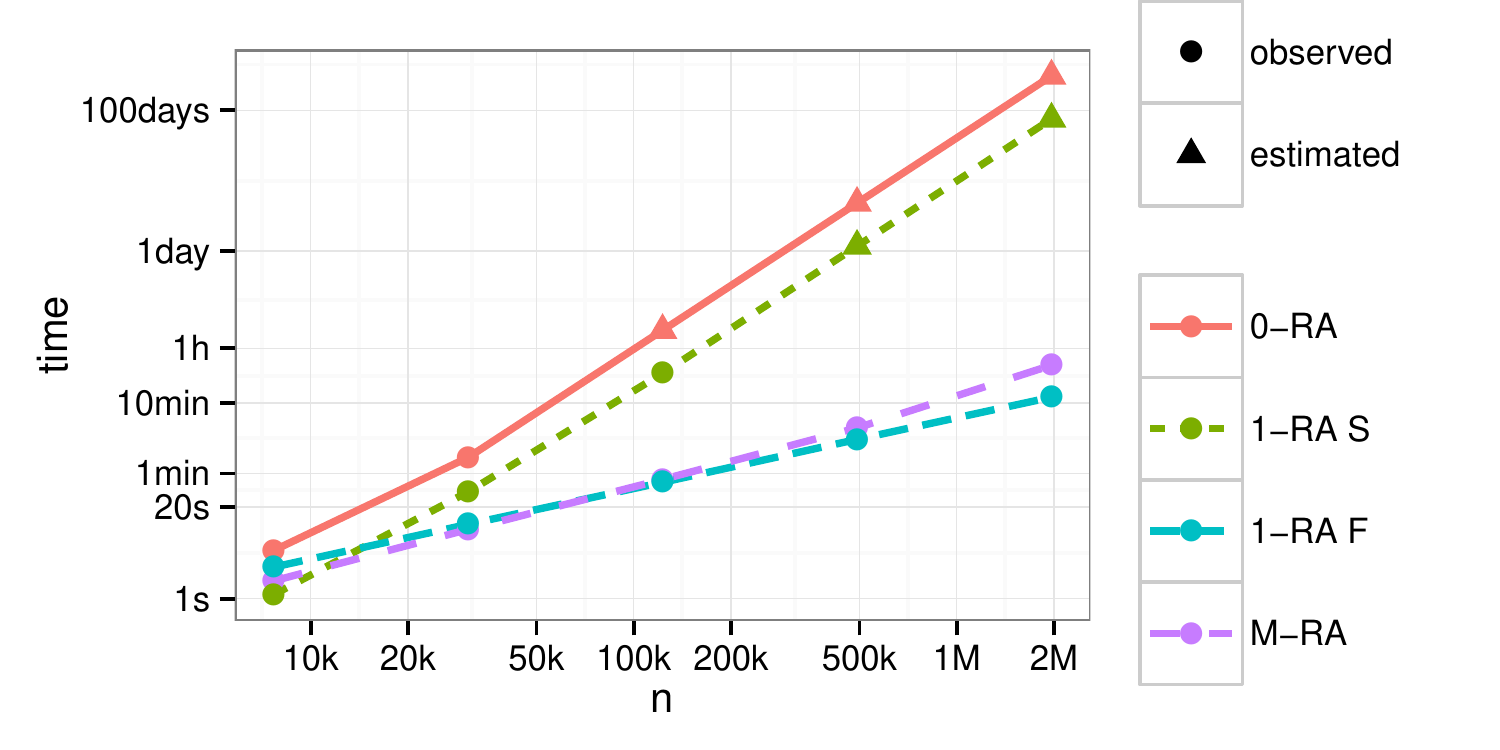}
	\caption{Times for likelihood evaluations}
	\label{fig:simtimes}
	\end{subfigure}
\hfill
	\begin{subfigure}{.48\textwidth}
	\centering
	\includegraphics[trim = 0mm 3mm 0mm 0mm, clip, width =1\linewidth]{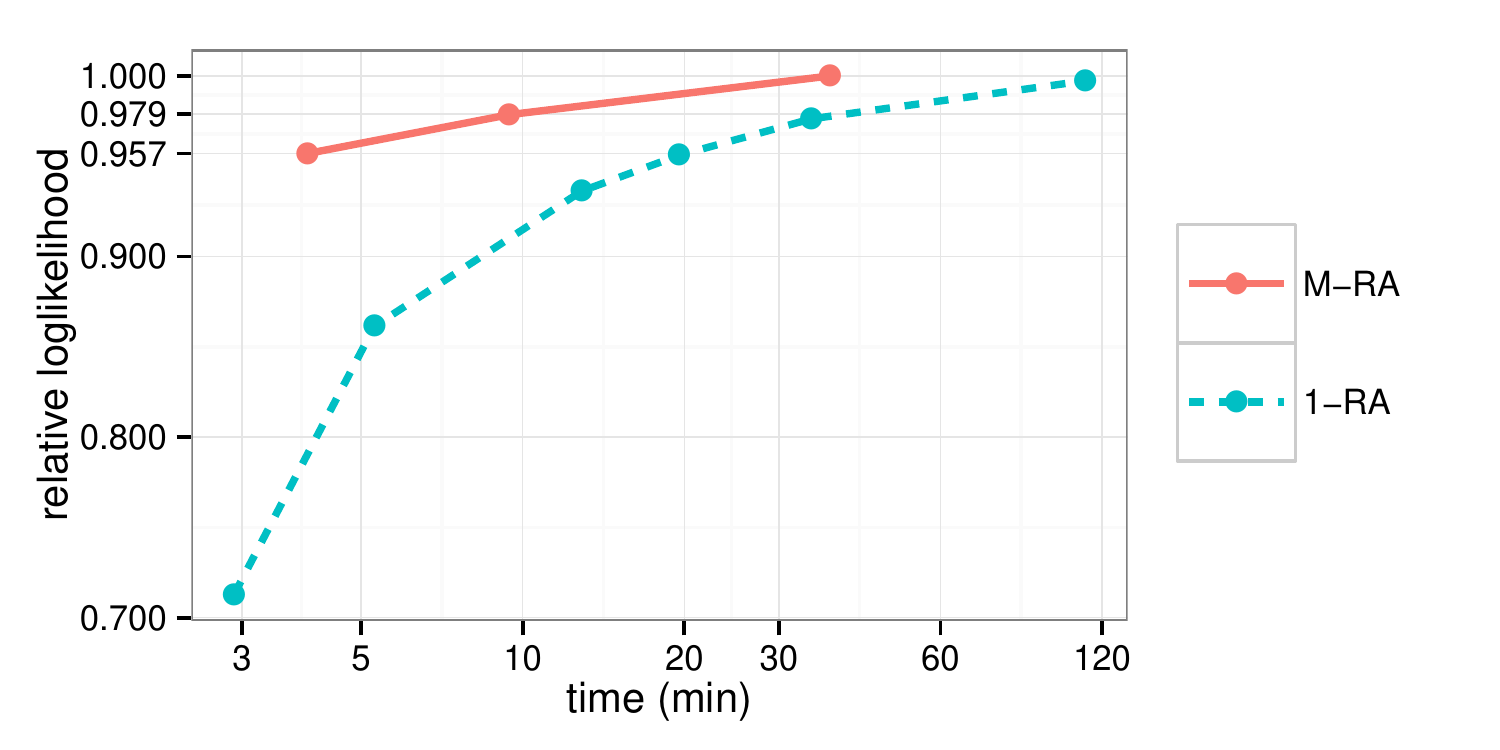}
	\caption{Comparison for fixed $n$=1,966,080}
	\label{fig:simfixedn}
	\end{subfigure}%
\end{subfigure}
\caption{Results from the simulation study in one spatial dimension described in Section \ref{sec:simstudy}. $0$-RA is the true Gaussian process, and 1-RA S and 1-RA F are full-scale approximations with increasing and fixed $r$, respectively. Note that all axes indicating time or sample size are on a log scale. The log-scores (i.e., loglikelihoods) are all scaled relative to the log-score of the $M$-RA.}
\label{fig:simresults}
\end{figure}

We then recorded the loglikelihood and the time taken to compute it for each $n$ and for each of the following four competitors:
\begin{description}[itemsep=0mm,topsep=1mm]
\item[0-RA:] A Gaussian process with the true covariance function $C_0$, which provides the best possible fit, but scales as $\order(n^3)$.
\item[1-RA F:] A ``fast'' 1-RA with fixed $r=240$ and increasing $J=n/240$, which scales as $\order(n)$.
\item[1-RA S:] A ``slow'' 1-RA with fixed $J=64$ and increasing $r=n/64$, which scales as $\order(n^3)$.
\item[M-RA:] A $M$-RA as described in Section \ref{sec:complexity}, with $r=30$, $J=4$ and $M=\log_4(n/30)$, which scales as $\order(n\log^2 n)$.
\end{description}
For all competitors, the true covariance function (including all parameters) is assumed known, and we use the loglikelihood (at the true parameters) implied by each competitor as a measure of how well that competitor approximates this true covariance. The loglikelihood is equivalent to the log-score, which is a strictly proper scoring rule in the sense that it is maximized in expectation by the true model \citep[e.g.,][]{Gneiting2014}. This means that, on average, the $0$-RA will have the highest possible log-score.

The results of these experiments with increasing sample size (averaged over the five datasets) are shown in Panels (\subref{fig:simfixed})--(\subref{fig:simtimes}) of Figure \ref{fig:simresults}. The computation times scale roughly as expected. We extrapolated the computation times of the $0$-RA and 1-RA S for values of $n$ for which the simulation machine ran out of memory, but were able to compute the exact loglikelihoods for the $0$-RA using the Durbin-Levinson algorithm for up to $n\approx$ 500,000. The $M$-RA and the 1-RA F had similar computation times, with the latter becoming slightly faster for very large $n$. The log-scores of the $M$-RA appear to be getting closer to those of the (computationally infeasible) $0$-RA and 1-RA S with increasing $n$, while the log-scores of the 1-RA F become increasingly worse relative to the optimum.

For the largest sample size considered ($n= $ 1,966,080), we further investigated computation times and log-scores for different versions of the $M$-RA (with $r=30$ and $M=2,4,8$) and the 1-RA (with $r$ between 60 and 960). The results are shown in Panel (\subref{fig:simfixedn}) of Figure \ref{fig:simresults}. The 2-RA and the 4-RA were roughly 8.7 and 11.8, respectively, faster than the fastest 1-RA with an equal or greater log-score. None of the 1-RAs achieved a log-score as high as the 8-RA. 

\begin{figure}
\begin{subfigure}{1\textwidth}
	\begin{subfigure}{.48\textwidth}
	\centering
	\includegraphics[trim = 0mm 3mm 0mm 0mm, clip, width =1\linewidth]{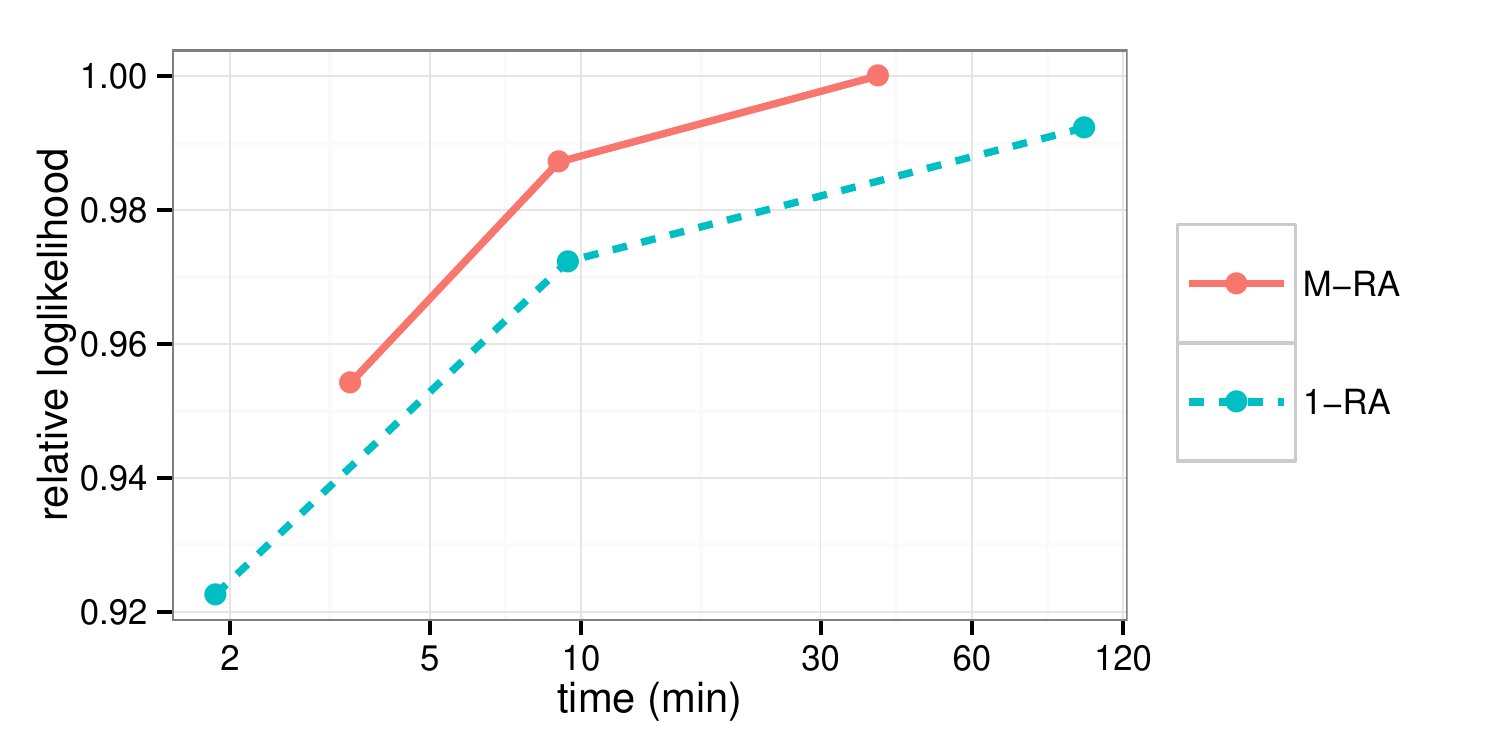}
	\caption{Without nugget}
	\label{fig:sim2D_nonug}
	\end{subfigure}
\hfill
	\begin{subfigure}{.48\textwidth}
	\centering
	\includegraphics[trim = 0mm 3mm 0mm 0mm, clip, width =1\linewidth]{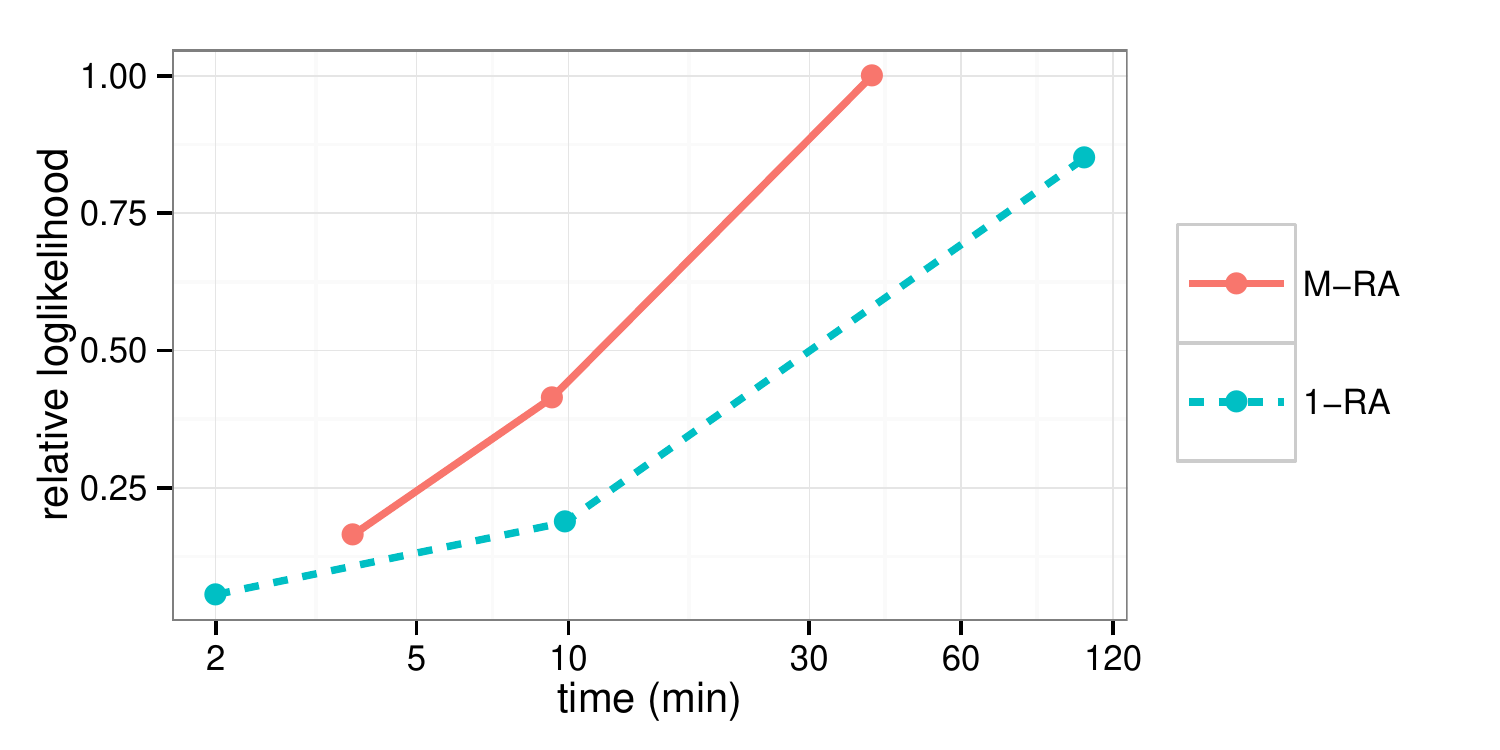}
	\caption{With nugget}
	\label{fig:sim2D_nug}
	\end{subfigure}%
\end{subfigure}
\caption{For the simulation study in two dimensions, comparison of the $M$-RA (with varying $M$) to the 1-RA (with varying $r$). Note that the time axes are on a log scale. The loglikelihoods are all scaled relative to the loglikelihood of the 8-RA.}
\label{fig:simresults2D}
\end{figure}

We also conducted a simulation study in two-dimensional space. We considered $n= $ 3,211,264 observations with an exponential covariance function with scale 0.3 and variance 1 on a regular grid on the unit square $\domain = [0,1]^2$. Using the function \texttt{RFsimulate} in the R package \texttt{RandomFields} \citep{Schlather2015}, we simulated five datasets without a nugget, and five datasets with a nugget consisting of Gaussian white noise with variance 0.05. For both settings, we compared the $M$-RA with $r=49$ and $M=2,4,8$ to the 1-RA with $r$ between 49 and 784. The averaged results are shown in Figure \ref{fig:simresults2D}. The relative performance of the two methods is similar to the one-dimensional case in Figure \ref{fig:simresults}(\subref{fig:simfixedn}), but in the case without a nugget even fast approximations with small $r$ or $M$ achieve a relatively high loglikelihood.


\subsection{Analysis of total precipitable water \label{sec:tpw}}

\begin{figure}
\flushleft
	\begin{subfigure}{.5\textwidth}
	\centering 
	\includegraphics[trim = 0mm 23mm 0mm 0mm, clip, width =.75\linewidth]{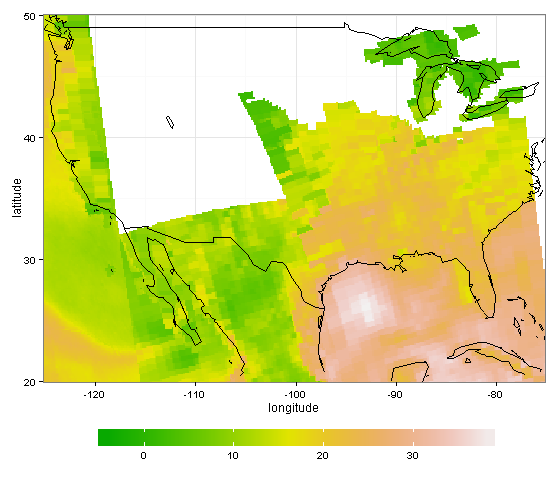}
	\caption{TPW measurements}
	\label{fig:tpwdata}
	\end{subfigure}
\begin{subfigure}{1\textwidth}
\centering
	\begin{subfigure}{.5\textwidth}
	\centering
	\includegraphics[trim = 0mm 23mm 0mm 0mm, clip, width =.75\linewidth]{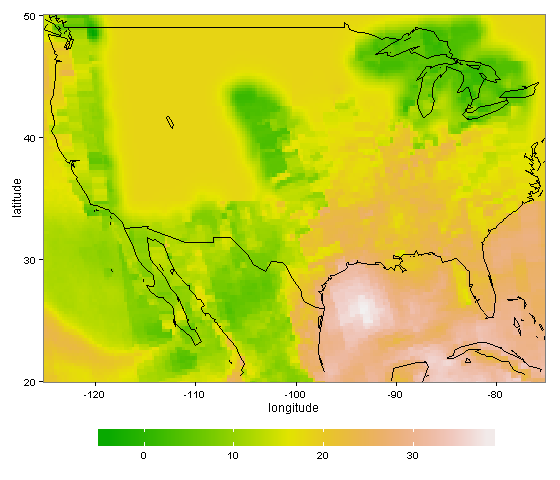}
	\caption{Posterior means for the 6-RA}
	\label{fig:tpwpred6}
	\end{subfigure}%
	\begin{subfigure}{.5\textwidth}
	\centering
	\includegraphics[trim = 0mm 23mm 0mm 0mm, clip, width =.75\linewidth]{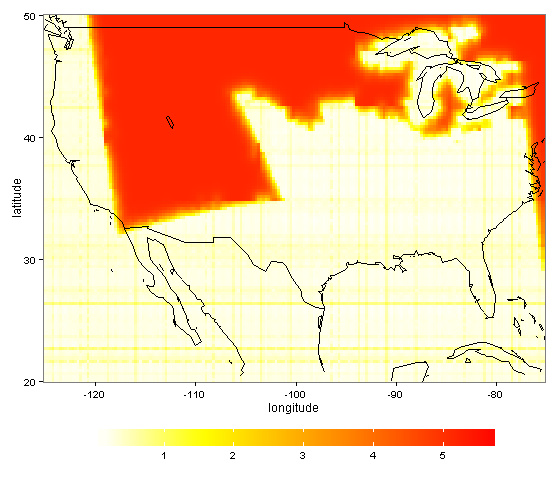}
	\caption{Posterior standard deviations for the 6-RA}
	\label{fig:tpwsd6}
	\end{subfigure}%
\end{subfigure}
\begin{subfigure}{1\textwidth}
\centering
	\begin{subfigure}{.5\textwidth}
	\centering
	\includegraphics[trim = 0mm 23mm 0mm 0mm, clip, width =.75\linewidth]{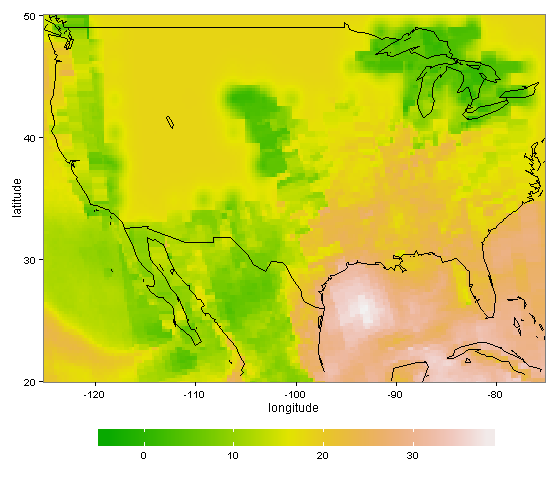}
	\caption{Posterior means for the 1-RA}
	\label{fig:tpwpred6}
	\end{subfigure}%
	\begin{subfigure}{.5\textwidth}
	\centering
	\includegraphics[trim = 0mm 23mm 0mm 0mm, clip, width =.75\linewidth]{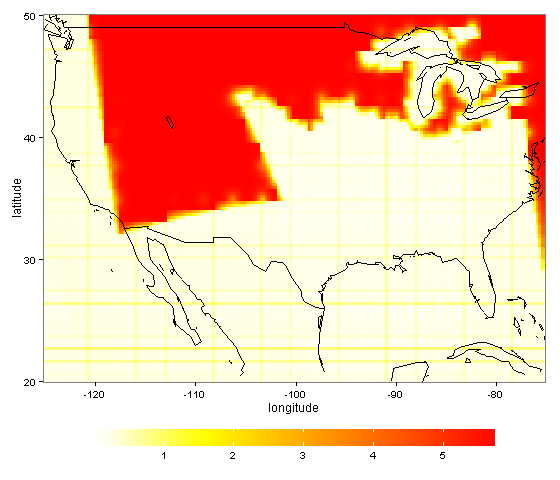}
	\caption{Posterior standard deviations for the 1-RA}
	\label{fig:tpwsd6}
	\end{subfigure}%
\end{subfigure}
\begin{subfigure}{1\textwidth}
\centering
	\begin{subfigure}{.5\textwidth}
	\centering
	\includegraphics[trim = 0mm 3mm 0mm 0mm, clip, width =.75\linewidth]{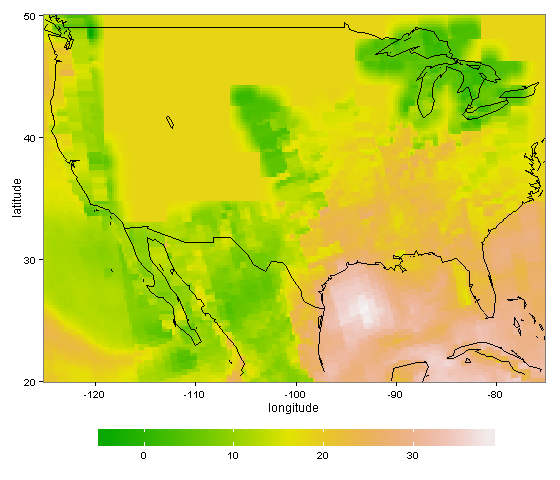}
	\caption{Posterior means for the block approx.}
	\label{fig:tpwpred6}
	\end{subfigure}%
	\begin{subfigure}{.5\textwidth}
	\centering
	\includegraphics[trim = 0mm 3mm 0mm 0mm, clip, width =.75\linewidth]{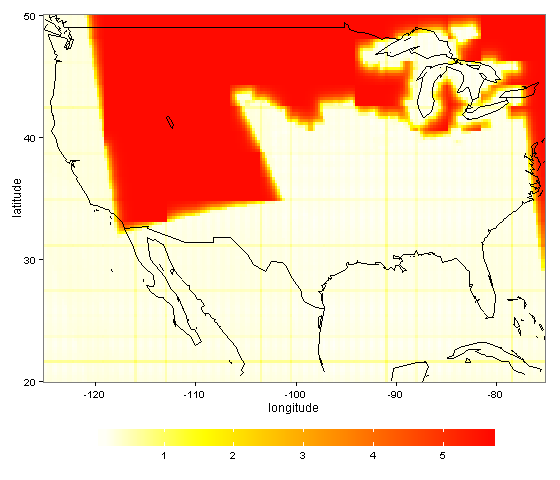}
	\caption{Posterior standard deviations for the block approx.}
	\label{fig:tpwsd6}
	\end{subfigure}%
\end{subfigure}
\caption{271,014 measurements of total precipitable water (TPW), along with posterior predictive means and standard deviations of the true underlying TPW field on a $0.25\dg \times 0.25\dg$ grid using three different methods. Color scales are in units of mm.}
\label{fig:tpw}
\end{figure}

We also applied our methodology to $n =$ 271,014 measurements of total precipitable water (TPW) made by the Microwave Integrated Retrieval System (MIRS) satellites between 2am and 3am UTC on February 1, 2011, over a region covering the United States.  The measurements are shown in Panel (\subref{fig:tpwdata}) of Figure \ref{fig:tpw}. The data are noisy and hourly datasets exhibit large gaps, which means that prediction of the true underlying TPW field is necessary at both observed and unobserved locations. Currently, an ad-hoc operational version of such a gap-filled product is sent to National Weather Service offices, where it is used to track the movement of water vapor in the atmosphere and to detect conditions that can lead to heavy precipitation (see \citealp{Kidder2007}, and \citealp{Forsythe2012}, for more details).

We extended our methodology slightly to accommodate the fact that the TPW observations contain measurement error. We assumed that the observations were
\[
z(\bs_i) = y_M(\bs_i) + \epsilon(\bs_i), \quad i=1,\ldots,n,
\]
where $y_M(\cdot)$ is the $M$-RA as before, and for simplicity we assumed that we have spatially independent measurement error, $\epsilon(\bs_i) \stackrel{iid}{\sim} N(0,\sigma^2_\epsilon)$. In this case, parameter inference and prediction (of $y(\cdot)$, not $z(\cdot)$) can proceed as before, except that we needed to set $\bfSigma_\jM = v_M(\locs_\jM,\locs_\jM) + \sigma^2_\epsilon \bI$ below \eqref{eq:matrices}.

We compared the proposed $M$-RA (with $M=6$) to the 1-RA (full-scale approximation) and to a block-independent approximation \citep[e.g.,][]{Stein2013a}, which simply divides the domain into subregions and treats the process as independent between subregions. This can be viewed as a special case of the 1-RA with zero knots at resolution $m=0$. The 6-RA had varying $J_m$ at different resolutions $m$, $(J_1,\ldots,J_6) = (2,2,4,8,8,16)$, with an average number of 16.45 knots per region. The 1-RA had 1,024 subregions with an average of 264.59 knots per region, and the block approximation had 256 subregions with an average of 1054.53 observations per region.
After subtracting a constant mean, some exploratory analyses showed that a Mat\'{e}rn covariance with smoothness parameter 1.5 fit the data well, and so all methods used were approximating a covariance of the form,
\[
C_0(\bs_1,\bs_2) = \sigma^2 \, \mathcal{M}_{1.5}(\|\bs_1-\bs_2\|/\kappa),
\]
where $\mathcal{M}_{1.5}$ is given in \eqref{eq:mat15}. 

We first estimated the unknown parameters $\sigma^2$, $\kappa$, and $\sigma^2_\epsilon$ by numerically maximizing the loglikelihood functions of the three approximation methods, and the resulting estimates and maximum loglikelihood values are given in Table \ref{tab:tpwresults}. Then, using the estimated parameters, we computed the posterior distribution of the underlying TPW field on a regular $0.25\dg \times 0.25\dg$ latitude/longitude grid of size 24,805 over the domain. Marginal summaries (posterior means and posterior standard deviations) are shown in Figure \ref{fig:tpw}. 

\begin{table}
\small
\begin{tabular}{lr|r|r|r|r|r|r|r|r|r}
                         &              & \multicolumn{5}{c|}{maximum likelihood estimation}                                                          & \multicolumn{2}{c|}{random} & \multicolumn{2}{c}{test regions} \\
 \multicolumn{1}{l|}{}                                        & $\bar r$  & time/lik.   & $\hat{\sigma}^2$ & $\hat\kappa$ & $\hat\sigma^2_\epsilon$ &  loglik. & RMSPE & CRPS         & RMSPE            & CRPS            \\ \hline
\multicolumn{1}{l|}{6-RA}   & 16.45      & 98.32     & 26.31                  & 0.74             & 0.29	                         & 1.00  & 0.56	   & 0.22	& 3.24	& 1.72       \\
\multicolumn{1}{l|}{1-RA}   & 264.59     & 108.05    & 33.21            & 0.78 &	0.29 & 	1.01	& 0.56	& 0.23	& 4.06	& 2.19 \\   
\multicolumn{1}{l|}{block}      & 1054.53	& 94.61 & 	31.22	& 0.73	& 0.29	& 1.00	& 0.56	& 0.23	& 4.51	& 2.38 \\
\multicolumn{1}{l|}{local}       &         	&         & 	 	       &  	      &  	       &  	       & 0.56	& 0.22	& 3.79	& 2.00
\end{tabular}
\caption{\label{tab:tpwresults} Results of the TPW analysis. block: block-independent approximation; local: local kriging based on 20 nearest neighbors; $\bar r$: average number of knots per region; time/lik.: average time (in seconds) per likelihood evaluation; loglik.: loglikelihood relative to the 6-RA loglikelihood; random: test set randomly sampled from observations; test regions: randomly selected test regions of size $5^\circ \times 5^\circ$; RMSPE: root mean-square prediction error; CRPS: mean continuous rank probability score (lower is better).}
\end{table}

Finally, we compared the posterior predictive distributions for three sets of 5,000 randomly selected held-out test data (to evaluate short-range predictions) and for three randomly selected held-out test regions of size $5^\circ \times 5^\circ$ (to evaluate long-range predictions). As the true TPW values are unknown, we compared the predictions to the observations, considering the mean-square prediction error (RMSPE) and the mean continuous rank probability score (CRPS). The CRPS is a strictly proper scoring rule that quantifies the fit of the entire predictive distribution (i.e., for a normal distribution, the mean and the variance) to the data, and it is on the same scale as the observations \citep[see, e.g.,][]{Gneiting2014}. 

As the block-independent long-range predictions were poor, we also carried out local kriging using the parameter estimates from the block-independent approximation. For every prediction location, the local-kriging predictions were based only on the data at the 20 nearest observed locations. The computation times for each test set were between 150 and 550 seconds. Note also that the $M$-RA methods provide the joint posterior predictive distribution at all prediction locations, while local kriging only provides marginal posterior predictive distributions at each prediction location.

Summarizing the comparison in Table \ref{tab:tpwresults}, the first three methods have similar computation times, maximum loglikelihood values, short-range predictions, and they all produce slight artifacts in the posterior-standard-deviation plots in the right column of Figure \ref{fig:tpw} in areas with nearly zero uncertainty. However, the 6-RA produces by far the best long-range predictions. In the prediction plots in the left column of Figure \ref{fig:tpw}, strong ``blocky'' artifacts are visible for both the 1-RA and the block approximation. These differences are important in many satellite-data applications, where large regions of missing data in hourly or daily data are very common due to satellite tracks and non-retrieval (e.g., because of heavy cloud coverage).

\section{Conclusions and Future Work \label{sec:conclusions}}

We have presented the multi-resolution approximation ($M$-RA), a novel technique for approximating Gaussian processes with any covariance function. The $M$-RA is essentially a linear combination of many spatial basis functions at multiple resolutions. The precision matrix of the basis-function weights has a multi-resolutional block-sparse structure, which allows scalable inference and distributed computations. Because the basis functions in our methodology are chosen optimally for a given covariance function, this can provide further insight on other multi-resolution approaches in which basis functions are chosen in a more ad-hoc way.

The $M$-RA compares favorably with the full-scale approximation of \citet{Sang2011a}, which is a current state-of-the-art method for large spatial data and can be viewed as a special case of the $M$-RA (with $M=1$). Using theoretical results, a toy example, large simulated datasets, and a real-data application, we have shown that the $M$-RA can provide a better approximation at the same computational complexity and computation time as the 1-RA, or it can provide a similar approximation at a fraction of the computational time. It should also be noted that our inference results for $M=1$ provide an algorithm for parallel, distributed computations for inference in the full-scale approximation.

We are planning on providing user-friendly software that provides good default choices for the $M$-RA and that can be run on both desktop computers and on high-performance computing environments. Taking advantage of the distributed-memory architecture of the latter should in principle allow applying the $M$-RA to datasets with hundreds of millions of observations, as many satellite instruments are now able to produce on a daily basis.

The $M$-RA not only approximates the data covariance matrix, but it is a valid Gaussian process in its own right. Extensions to more complicated scenarios are therefore possible by embedding the $M$-RA process in a hierarchical model \citep[e.g.,][]{Cressie2011}. When the data measurement process is complex, the $M$-RA can be embedded in a hierarchical model that explicitly models the measurement process, and allows, for example, modeling non-Gaussian data, or fusing data from different measurement instruments. 

Also of interest is a spatio-temporal version of the $M$-RA. Because it is possible to store and propagate the entire joint posterior predictive distribution, the $M$-RA could be extended to allow Kalman-filter-type inference in massive spatio-temporal state-space models \citep[which is challenging for other sparse-precision approaches such as][]{Lindgren2011a}. In this sense, the $M$-RA might also provide an alternative to the ensemble Kalman filter \citep{Evensen1994,Katzfuss2015b} in certain situations.


\footnotesize
\appendix
\section*{Acknowledgments}
This research was partially supported by NASA's Earth Science Technology Office AIST-14 program and by National Science Foundation (NSF) Grant DMS-1521676.
I would like to thank Dorit Hammerling, Doug Nychka, Mikyoung Jun, Joe Guinness, Huiyan Sang, Suhasini Subba Rao, Valen Johnson, and two anonymous referees for helpful comments and suggestions. I am also grateful to John Forsythe and Stan Kidder for providing the dataset in Section \ref{sec:tpw} and helpful advice.

\section{Proofs \label{app:proofs}}

\begin{proof}[\bfseries Proof of Proposition \ref{prop:nnd}]

For any set of locations $\locs \subset \domain$, $\by_M(\locs)$ in the form \eqref{eq:mradef} is a linear combination of the vector consisting of all basis-function weights, which has a  multivariate normal distribution, and so $y_M(\cdot)$ is a Gaussian process.

Further, note that $y_M(\cdot)$ in \eqref{eq:mra1} is a sum of independent components, $\pp_m(\bs)= E\big(\delta_m(\bs) | \bfdelta_m(\knots^{(m)}) \big)$, where $\delta_m(\cdot)$ is independent between regions $\domain_\jm$. Starting with $\delta_0(\cdot) = y_0(\cdot)$, we can show iteratively for $m=1,\ldots,M-1$ using the law of total variance that, for any finite set $\locs_\jm \subset \domain_\jm$, the matrix
\begin{align*}
\var\big(\bfdelta_m(\locs_\jm)\big) &= \var\big(\bfdelta_{m-1}(\locs_\jm)\big) - \var\big(E\big(\bfdelta_{m-1}(\locs_\jm)|\bfdelta_{m-1}(\knots_\jmm)\big)\big) \\
   &= \var\big(\bfdelta_{m-1}(\locs_\jm)|\bfdelta_{m-1}(\knots_\jmm)\big)
\end{align*}
is nonnegative definite. Thus, the covariance functions of the $\delta_m(\cdot)$, the $\pp_m(\cdot)$, and of $y_M(\cdot)$ are nonnegative definite.
\end{proof}

\begin{proof}[\bfseries Proof of Proposition \ref{prop:prediction}]

For $\jm=1,\ldots,J$, $m=0,1,\ldots,M$, and $l=0,\ldots,m$, define
\begin{align*}
 \bfmu_\jm^l & \colonequals E\big(\by_M(\locs^P_\jm) | \by_M(\locs),\etaset_{l-1}\big),\\
 \bfPsi_{\im;\jm}^l & \colonequals \cov\big(\by_M(\locs^P_\im),\by_M(\locs^P_\jm) | \by_M(\locs),\etaset_{l-1}\big),\\
 \bB_\jm^{l,P} & \colonequals \bb_\jl(\locs^P_\jm)\\
 \bL_{\jm}^l & \colonequals \cov\big( \by_M(\locs^P_\jm), \by_M(\locs_\jl) \big| \etaset_{l-1} \big) = v_l(\locs^P_\jm,\locs_\jl),\\
\widetilde{\bB}^{l,k}_\jm & \colonequals \bB_\jm^{k,P} - \bL_\jm^l \bfSigma^{-1}_\jl \bB^k_\jl, \quad k=0,\ldots,l-1.
\end{align*}

Note that, for $l<M$, we have $\bL_{\jm}^l = \bB_\jm^{l,P} \bK_\jl \bB_\jl^l{}'+\widetilde{\bL}_{\jm}^l$, where $\widetilde{\bL}_{\jm}^l$ is a sparse block matrix with the only nonzero block being $\bL_\jm^{l+1}$. Hence, it can be shown that $\widetilde{\bL}_{\jm}^l\bV^{-1}_\jl \bB^k_\jl = \bL^{l+1}_\jm \bfSigma^{-1}_\jlp\bB^k_\jlp$. Using a variant of the Sherman-Morrison-Woodbury formula, we can also show that $\bK_\jl \widetilde{\bA}_\jl^{l,k} =\widetilde{\bK}_\jl \bA_\jl^{l,k}$. By applying \eqref{eq:smw}, we therefore have
\begin{align*}
\widetilde{\bB}^{l,k}_\jm & = \bB^{k,P}_\jm - \bB_\jm^{l,P} \bK_\jl (\bB_\jl^l{}'\bV_\jl^{-1}\bB_\jl^k \\
    & \qquad \qquad - \bB_\jl^l{}'\bV_\jl^{-1}\bB_\jl^l \widetilde{\bK}_\jl\bB_\jl^l{}'\bV_\jl^{-1}\bB_\jl^k) \\
  & \qquad - \widetilde{\bL}^l_\jm \bV_\jl^{-1}\bB_\jl^k + \widetilde{\bL}^l_\jm \bV_\jl^{-1}\bB_\jl^l \widetilde{\bK}_\jl \bA_\jl^{l,k} \\
& = \bB^{k,P}_\jm - \bB_\jm^{l,P} \bK_\jl (\bA_\jl^{l,k} - \bA_\jl^{l,l} \widetilde{\bK}_\jl \bA_\jl^{l,k})\\
  & \qquad -\bL^{l+1}_\jm \bfSigma^{-1}_\jlp\bB^k_\jlp + \bL^{l+1}_\jm \bfSigma^{-1}_\jlp\bB^l_\jlp \widetilde{\bK}_\jl \bA_\jl^{l,k}\\
& = \bB^{k,P}_\jm -\bL^{l+1}_\jm \bfSigma^{-1}_\jlp\bB^k_\jlp \\
  & \qquad - (\bB_\jm^{l,P} - \bL^{l+1}_\jm \bfSigma^{-1}_\jlp\bB^l_\jlp)\widetilde{\bK}_\jl \bA_\jl^{l,k}\\
& = \widetilde{\bB}^{l+1,k}_\jm - \widetilde{\bB}^{l+1,l}_\jm \widetilde{\bK}_\jl \bA^{l,k}_\jl 
\end{align*}
which proves \eqref{eq:bfrecursion}.

It is easy to see that the desired posterior predictive distribution is multivariate normal, $\by_M(\locs^P) | \by_M(\locs)\sim N(\bfmu,\bfPsi)$, and so spatial prediction amounts to finding the posterior mean and covariance matrix, $\bfmu$ and $\bfPsi$, respectively. To obtain these quantities, note that we have from well-known properties of the multivariate normal distribution that
\begin{equation}
\label{eq:predmean}
\begin{split}
 \bfmu_\jm^l & \textstyle = E\big(\by_M(\locs^P_\jm) | \by_M(\locs_\jl),\etaset_{l-1}\big)\\
   & \textstyle = \sum_{k=0}^{l-1} \bB_\jm^{k,P} \bfeta_\jk + \bL^l_\jm \bfSigma_\jl^{-1}\big( \by_M(\locs_\jl) -  \sum_{k=0}^{l-1} \bB_\jm^{k} \bfeta_\jk \big) \\
   & \textstyle  = \bL^l_\jm \bfSigma_\jl^{-1} \by_M(\locs_\jl) + \sum_{k=0}^{l-1} \widetilde{\bB}_\jm^{l,k} \bfeta_\jk .
\end{split}
\end{equation}
By the law of total expectation, we therefore have
\begin{equation}
\label{eq:predmeanrec}
\begin{split}
 \bfmu_\jm^{l-1} & \textstyle  = E\big(  \bfmu_\jm^l \big| \by_M(\locs), \etaset_{l-2} \big) \\
    & \textstyle = \bL^l_\jm \bfSigma_\jl^{-1} \by_M(\locs_\jl) +\sum_{k=0}^{l-2} \widetilde{\bB}_\jm^{l,k} \bfeta_\jk + \widetilde{\bB}_\jm^{l,l-1} \widetilde{\bfnu}_{j_1,\ldots,j_{l-1}}\\
    & \textstyle = \bL^l_\jm \bfSigma_\jl^{-1} \by_M(\locs_\jl) + \widetilde{\bB}_\jm^{l,l-1} \widetilde{\bK}_{j_1,\ldots,j_{l-1}} \widetilde{\bfomega}_{j_1,\ldots,j_{l-1}} +\sum_{k=0}^{l-2} \widetilde{\bB}_\jm^{l,k} \bfeta_\jk,
\end{split}
\end{equation}
and by the law of total covariance, we have
\begin{equation}
\label{eq:predcovrec}
\begin{split}
 \bfPsi_{\im;\jm}^{l-1} & \textstyle = E\big(  \bfPsi_{\im;\jm}^l \big| \by_M(\locs), \etaset_{l-2} \big) + \cov\big( \bfmu_\im^l, \bfmu_\jm^l \big| \by_M(\locs), \etaset_{l-2} \big) \\
    & \textstyle = \bfPsi_{\im;\jm}^l + \cov\big( \widetilde{\bB}_\im^{l,l-1} \bfeta_{i_1,\ldots,i_{l-1}}, \widetilde{\bB}_\jm^{l,l-1} \bfeta_{j_1,\ldots,j_{l-1}} \big| \by_M(\locs), \etaset_{l-2} \big) \\
    & \textstyle = \bfPsi_{\im;\jm}^l + \widetilde{\bB}_\jm^{l,l-1} \widetilde{\bK}_{j_1,\ldots,j_{l-1}} \widetilde{\bB}_\jm^{l,l-1}{}'\, I((\im) = (\jm)) .
\end{split}
\end{equation}

The result \eqref{eq:prediction} follows by starting with $\bfmu_\jM^M$ from \eqref{eq:predmean} and $\bfPsi_\jM^M = \bV^P_\jM - \bL_{\jM}^M\bfSigma^{-1}_\jM\bL_{\jM}^M{}'$, and iteratively applying \eqref{eq:predmeanrec} and \eqref{eq:predcovrec} with $m=M$ for $l=M,\ldots,0$.

\end{proof}

\footnotesize
\bibliographystyle{apalike}
\bibliography{library}

\end{document}